\documentclass[aip,pop,reprint]{revtex4-1}
 
\newcommand{\D}{\textrm{D}}
\newcommand{\B}{\textrm{B}}

\newcommand{\be}{\begin{eqnarray}}
\newcommand{\ee}{\end{eqnarray}}
\newcommand{\ben}{\begin{eqnarray*}}
\newcommand{\een}{\end{eqnarray*}}

\newcommand{\vc}{\mathbf}
\usepackage{graphicx}
\usepackage{color}
\usepackage{amsmath,amssymb}

\begin{document}


\title{Extended space and time correlations in strongly magnetized plasmas}


\author{Keith R.~Vidal}
\affiliation{Department of Physics and Astronomy, University of Iowa, Iowa City, Iowa 52242, USA}
\author{Scott D.~Baalrud}
\email{baalrud@umich.edu}
\affiliation{Department of Nuclear Engineering and Radiological Sciences, University of Michigan, Ann Arbor, MI 48109, USA}


\date{\today}

\begin{abstract}

Molecular dynamics simulations are used to show that strong magnetization significantly increases the space and time scales associated with interparticle correlations.
The physical mechanism responsible is a channeling effect whereby particles are confined to move along narrow cylinders with a width characterized by the gyroradius and a length characterized by the collision mean free path. 
The predominant interaction is $180^\circ$ collisions at the ends of the collision cylinders, resulting in a long-range correlation parallel to the magnetic field. 
Its influence is demonstrated via the dependence of the velocity autocorrelation functions and self-diffusion coefficients on the domain size and run time in simulations of the one-component plasma. 
A very large number of particles, and therefore domain size, must be used to resolve the long-range correlations, suggesting that the number of charged particles in the collection must increase in order to constitute a plasma. 
Correspondingly, this effect significantly delays the time it takes to reach a diffusive regime, in which the mean square displacement of particles increases linearly in time. 
This result presents challenges for connecting measurements in non-neutral and ultracold neutral plasma experiments, as well as molecular dynamics simulations, with fluid transport properties due to their finite size. 

\end{abstract}




\maketitle


\section{Introduction}

Plasmas are collections of charged particles large enough to exhibit collective behavior.\cite{Gurnett:Book} 
Correspondingly, the assumption of a scale separation between the short-range interactions responsible for collective motion at the microscopic scale, and the longer-range macroscopic evolution is inherent in the development of plasma kinetic theory.
This scale is commonly characterized by the Debye length because shielding limits the range over which particles interact. 
Thus, the collection of charges must be much larger than the Debye length in order to satisfy the definition of a plasma.
However, this expectation is based on the assumption that the plasma is weakly magnetized in the sense that gyromotion is negligible at the microscopic correlation scale. 

In this work, we show that the microscopic correlation scale characterizing interactions between particles can increase substantially if the plasma is strongly magnetized. 
In this context, strong magnetization refers to plasmas in which the typical gyrofrequency of particles significantly exceeds the plasma frequency: $\beta = \omega_c/\omega_p \gg 1$, where $\omega_c = qB/m$ and $\omega_p = \sqrt{e^2 n/\epsilon_o m}$. 
Equivalently, this corresponds to particle gyroradii that are smaller than the Debye length, indicating that magnetization influences inter-particle collisions. 
In particular, our molecular dynamics (MD) simulations of the one-component plasma (OCP) show that strong magnetization causes particles to become confined to cylinders with a length that is characterized by a mean free path for 180$^\circ$ collisions, and a width that is characterized by the gyroradius. 
The length of these cylinders increases as $\beta$ increases. 
This long-range correlation dramatically extends the scale separating microscopic interactions from macroscopic behavior when the magnetization strength is large.
It has implications for connecting experimental measurements~\cite{Beck:1992,Beck:1996,Dubin:1999,HollmannPRL1999,KrieselPRL2001,AndereggPOP2017,AffolterPRL2016}, as well as MD simulations~\cite{Ott:2011,OttPRE2015,OttPRE2017,Baalrud:regimes,ScheinerPRE2020,BernsteinPRE2020}, to fluid transport processes.

Strongly magnetized plasmas arise in a number of contexts. 
Non-neutral plasmas, such as those used to trap antimatter,~\cite{Danielson:2015} can be very strongly magnetized $\beta \gg 10$.~\cite{Beck:1992,Beck:1996,Dubin:1999} 
Electrons in a number of other laboratory experiments, including ultracold neutral plasmas,~\cite{Roberts:2020} dusty plasmas,~\cite{Thomas:2012,Bonitz:2012,TadsenPRE2018,HartmannPRE2019} high-field inertial confinement fusion experiments,~\cite{Gotchev:2009,Hohenberger:2012,Gomez:2014} and even to some extent in conventional magnetic confinement fusion experiments,~\cite{Shimada:2007,Paz-Soldan:2014,Creely:2020} reach regimes in which $\beta > 1$.
In nature, strongly magnetized plasmas are encountered in several astrophysical contexts, including the magnetospheres of some planets and exoplanets,~\cite{Khurana:2004} and the atmosphere of neutron stars.~\cite{Harding:2006} 
Many of these examples are influenced by moderate or strong Coulomb coupling simultaneously with strong magnetization. 
Coulomb coupling can be quantified using the Coulomb coupling parameter, $\Gamma=(q^2/a)/(4\pi \epsilon_o k_\B T)$, which is the ratio of the Coulomb potential energy at the average interparticle spacing, $a=(3/4 \pi n)^{1/3}$, to the average kinetic energy. 
Understanding how the combined effects of strong magnetization and strong coupling influence transport, and the system scale required to exhibit a plasma regime, is important to developing models to describe these systems. 

Here, we show how coupling and magnetization influence the space and time requirements for reaching a plasma regime using MD simulations of the OCP. 
The space requirement was determined through convergence tests of the parallel, perpendicular, and transverse velocity autocorrelation functions ($Z_\parallel$, $Z_\perp$ and $Z_\wedge$), and self-diffusion coefficients ($D_\parallel$, $D_\perp$ and $D_\wedge$), with respect to the size of the simulation domain at a fixed temperature and density. 
In the regime of weak coupling ($\Gamma \ll 1$) and weak magnetization ($\beta < 1$), results agree with the conventional expectation that the domain size must exceed the Debye length $\lambda_D= \sqrt{\epsilon_o k_\B T / (q^2 n)}$.~\cite{Hockney:MD} 
However, if the plasma is weak to moderately coupled ($\Gamma \lesssim 1$) but strongly magnetized ($\beta \gg 1$), the domain size requirement becomes much larger and increases with increasing magnetic field strength. 
This directly demonstrates a long-range correlation effect that extends the scale of inter-particle interactions. 
At strong coupling ($\Gamma \gtrsim 1$) and weak magnetization ($\beta < 1$), the domain size requirement is observed to be determined by the inter-particle spacing, as expected. 
Strong magnetization is again observed to extend the spatial scale of these interactions, but to a lesser degree than at weak or moderate coupling. 

This result implies that using MD simulations to characterize strongly magnetized plasmas is fraught with challenges, as they require particle numbers that are very large (MD simulation are usually limited to $\sim 10^6$ particles due to computational expense).
It also implies that some experiments, such as non-neutral~\cite{Beck:1992,Beck:1996,HollmannPRL1999,KrieselPRL2001,AffolterPRL2016,AndereggPOP2017,Dubin:1999} and ultracold neutral plasmas,~\cite{TiwariPOP2018,Gorman:2020,Roberts:2020,IsaevJPB2018} may not be large enough to represent a three-dimensional plasma regime. 
These experiments access a wide range of particle numbers $10^4-10^8$, as well as coupling and magnetization strengths, but encounter the same situation as the simulations in that very large particle numbers are required to represent a plasma at high $\beta$. 
For example, particle numbers on the order of $10^6$, $\Gamma$ values near $\simeq 0.1$ and large $\beta$ values ($> 10$) are common in non-neutral plasma experiments.~\cite{Beck:1992,Beck:1996,HollmannPRL1999,KrieselPRL2001,AffolterPRL2016,AndereggPOP2017,Dubin:1999}

The extended scale of spatial correlations implies a corresponding extension of their timescale. 
Since the channeling effect generated by strong magnetization causes particles to travel much further in the parallel dimension before colliding with another particle, the time between these collisions increases as $\beta$ increases. 
This is demonstrated through the relaxation time of the velocity autocorrelation functions and the mean square displacement of particles from their initial positions in each the parallel, perpendicular and transverse directions. 
By the Einstein relation, the fluid process of diffusion corresponds to the regime in which the mean square displacement increases linearly in time.~\cite{Hansen:2013} 
The time it takes to reach this regime is related to the time it takes particles to interact, which is several plasma periods ($\omega_p^{-1}$) in a weakly magnetized plasma $(\beta \lesssim 1)$. 
Results show that strong magnetization significantly increases the time that it takes the mean square displacement to reach a linear in time regime, thus delaying the time it takes to reach the fluid limit. 

The delayed time to reach a fluid regime has implications for both MD simulations and experiments. 
Simulations become much more expensive because the timestep must be set to resolve the very fast gyromotion, while the total simulation time becomes longer because of the increased correlation time. 
Previous MD simulations of diffusion in strongly magnetized 2D Yukawa plasmas have reported superdiffusive behavior,~\cite{Feng:2014} whereby the mean square displacement in the perpendicular direction increases at a rate faster than proportional to time; i.e., as $t^\alpha$ where $\alpha > 1$.
This was later observed experimentally, but thought to be associated with not reaching the time asymptotic limit associated with the hydrodynamic regime~\cite{HartmannPRE2019}. 
We also observe this behavior in our 3D simulations when they are not run sufficiently long.
However, eventually, the mean square displacement reaches the normal diffusive regime in which it increases linearly in time.
This suggests that true diffusion persists in strongly magnetized plasmas in three-dimensions, but the time at which the diffusive regime is reached is delayed significantly. 
Experimentally, this has implications for the length of time over which the plasma must evolve, or measurements must be taken, in order to probe fluid properties. 
For example, ultracold neutral plasmas expand at a rate determined by ion dynamics, which could become comparable to the time it takes electrons to reach a fluid regime if they are sufficiently strongly magnetized. 

The remainder of this paper is organized as follows. 
Section~\ref{sec:setup} describes the MD simulations and the calculation of diffusion coefficients.  
Section~\ref{sec:unmagnetized} summarizes results for the correlations scales in unmagnetized plasmas as coupling strength varies.  
Section~\ref{sec:magnetized} shows how these space and time scales dramatically extend in response to an applied magnetic field. 
Conclusions are discussed in Sec.~\ref{sec:conclusions}. 

\section{Simulation setup\label{sec:setup}}

Simulations were prepared by randomly assigning positions and velocities to $N$ charged particles in a cubic box with periodic boundary conditions.  
Velocities and positions were updated by using the Velocity-Verlet algorithm.~\cite{Frenkel:MD}  
In simulations with an applied magnetic field a modified Velocity-Verlet algorithm provided by Spreiter and Walter~\cite{Spreiter:1999} that allows for an arbitrarily strong uniform external magnetic field was used.  
The simulation was performed in two stages.  
In the equilibrium stage the system was brought into constant energy configuration using a velocity rescaling thermostat.~\cite{Frenkel:MD}  
After equilibrium was achieved, particles evolved in time in the microcanonical ensemble during the main run stage.  
In simulations with an external magnetic field, the field was turned on only during the main run stage and was not present during the equilibrium stage in order to reduce computational time.  
The presence of an external magnetic field in a classical system does not change the equilibrium state via the Bohr-van-Leeuwen theorem,~\cite{Pathria:2011} thus turning on the magnetic field during the main run stage does not change the energy or configuration of the equilibrium state.

The Coulomb potential was implemented using an Ewald Summation technique through the Particle-Particle Particle-Mesh ($\text{P}^3 \text{M}$) algorithm.~\cite{Hockney:MD}  
This splits the Coulomb potential into short-range (Particle-Particle) and long-range (Particle-Mesh) components.  
This representation of the potential allows for high resolution of close interactions and a computationally efficient long-range force which is computed on a mesh using three-dimensional fast Fourier transforms.

Positions and velocities of a random subset of particles were recorded during the main run stage of the simulation, and used to compute the velocity autocorrelation function in each of three vector directions 
\begin{subequations}
\label{eq:Z}
\begin{align}
	Z_{\parallel}(t) &= \langle v_{z}(t) v_{z}(0) \rangle \\
	Z_{\bot}(t)&=\frac{1}{2} \bigg[ \langle v_{x}(t) v_{x}(0) \rangle +  \langle v_{y}(t) v_{y}(0) \rangle \bigg] \\
	Z_{\wedge}(t)&=\frac{1}{2} \bigg[ \langle v_{x}(t) v_{y}(0) \rangle -  \langle v_{y}(t) v_{x}(0) \rangle \bigg]
\end{align}
\end{subequations}
where $\langle \ldots \rangle$ denotes an average over the ensemble of saved trajectories and
the coordinate orientation is
\begin{subequations}
\begin{align}
\vc{r}_{\parallel} &= (\hat{\vc{z}} \cdot \vc{r}) \hat{\vc{z}} \\
\vc{r}_{\bot}  &= \vc{r} - \vc{r}_{\parallel} \\
\vc{r}_{\wedge}  &= \hat{\vc{z}} \times \vc{r} .
\end{align}
\end{subequations}
Here, $\vc{B} = B \hat{\vc{z}}$, and the symbols $\parallel$, $\bot$, and $\wedge$ refer to the parallel, perpendicular, and transverse directions. 
In the absence of a magnetic field, $Z(t)  = \frac{1}{3}\langle \vc{v}(t) \cdot \vc{v}(0)\rangle = \frac{1}{3} Z_\parallel(t) + \frac{2}{3} Z_\bot(t)$ and $Z_\wedge(t) = 0$.
Diffusion coefficients were computed from the velocity autocorrelation functions using the Green-Kubo relations~\cite{Frenkel:MD} 
\begin{subequations}
\label{eq:d_gk}
\begin{align} 
\label{Dpara:eq}
D_{\parallel} &= \frac{k_BT}{m} \int_{0}^{\infty} Z_{\parallel}(t) dt \\
\label{Dperp:eq}
D_{\bot} &= \frac{k_BT}{m} \int_{0}^{\infty} Z_{\bot}(t) dt \\
\label{Dwedge:eq}
D_{\wedge} &= \frac{k_BT}{m} \int_{0}^{\infty} Z_{\wedge}(t) dt .
\end{align}
\end{subequations}
In the absence of a magnetic field $D = \frac{k_BT}{m} \int_0^\infty Z(t) dt = \frac{1}{3} D_\parallel + \frac{2}{3} D_\perp$, and $D_\wedge = 0$. 
Since the simulations can only be run for a finite time, the upper limit in Eq.~(\ref{eq:d_gk}) is the simulation run time $t_\textrm{run}$, which must be sufficiently long for the integral to converge. 
Determining the length of time necessary for this to occur is a focus of this study. 

In establishing the run time necessary to reach a diffusive regime, it is also informative to consider Einstein's description relating the diffusion coefficients to the mean square displacement of a particle from its initial position
\begin{subequations}
\label{eq:d_msd}
\begin{align}
\label{MSDpara:eq}
D_{\parallel} &= \lim_{t \to \infty} \frac {\langle | z(t) - z(0) |^2 \rangle}{2 t} \\ \label{MSDperp:eq}
D_{\bot} &= \lim_{t \to \infty} \frac {\langle | x(t) - x(0) |^2 \rangle + \langle | y(t) - y(0) |^2 \rangle}{4 t} \\
\label{MSDwedge:eq}
D_{\wedge} &= \lim_{t \to \infty} \frac {\langle | x(t) - y(0) |^2 \rangle - \langle | y(t) - x(0) |^2 \rangle}{4 t}.
\end{align}
\end{subequations}
In the unmagnetized limit $D = \lim_{t\rightarrow \infty} \langle | \vc{r}(t) - \vc{r}(0) |^2 \rangle/(6t)$. 
In practice, the mean square displacement was computed not from the particle positions, but from their velocities using $\vc{r}(t) - \vc{r}(0) = \int_0^t \vc{v}(t^\prime)dt^\prime$, so $\langle | \vc{r}(t) - \vc{r}(0) |^2 \rangle = 2\int_0^t dt^\prime \int_0^{t^\prime} dt^{\prime \prime} \langle \vc{v}(t^\prime) \cdot \vc{v}(t^{\prime \prime}) \rangle$, which is directly related to the velocity autocorrelation function~\cite{Hansen:2013} 
\begin{equation}
    \langle | \vc{r}(t) - \vc{r}(0) |^2 \rangle = 6t \int_0^t \biggl(1 - \frac{s}{t} \biggr) Z(s) ds .
\end{equation}
Each of the vector components in the magnetized case were obtained from using the appropriate component of the velocity autocorrelation function in this expression. 

\begin{table}
\caption{Unmagnetized OCP Simulation Parameters}
\centering
\begin{tabular*}{0.475\textwidth}{c @{\extracolsep{\fill}}c c c}
\hline \hline
$\Gamma$ & $\delta t (\omega_p^{-1})$ & $t_{\text{run}}(\omega_p^{-1})$ \\
\hline
$\geq 0.5$ & $0.01$ & $1638.40$ \\
$0.25-0.45$ & $0.0025$ & $1638.40$ \\
$0.075-0.225$ & $10^{-3}$ & $3276.80$ \\
$0.01-0.05$ & $10^{-4}$ & $8192$ \\
$10^{-3}$ & $10^{-5}$ & $1638.40$ \\
$10^{-4}$ & $10^{-6}$ & $102.40$ \\
\hline \hline
\end{tabular*}
\label{table:Unmag}
\end{table}    

The time-step, $\delta t$, and simulation duration, $t_{\text{run}}$, used for simulations of the unmagnetized OCP are shown in Table \ref{table:Unmag}.  
The time step was chosen in order to conserve energy to $\lesssim 10^{-5}$.  
Nominally, the simulation duration must be much greater than the decay time of the velocity autocorrelation functions, and converge at a rate faster than $t^{(-3/2)}$.~\cite{Alder:1970,Hansen:2013}  
This condition was satisfied for simulations with $\Gamma \geq 0.05$, but for those with $\Gamma < 0.05$ it was computationally impractical to achieve the necessary simulation run time. 
Instead, an alternative method based on a fit to a relaxation time approximation was applied (see Sec.~\ref{sec:unmagnetized} for details). 

Table \ref{table:Mag} shows a range of run times used in simulations with a magnetic field. 
For these, if $\beta \leq 1$ the time-step was chosen to be the same as in the absence of a magnetic field, but if $\beta > 1$ it was decreased to $\delta t = 0.01 \omega_{c}^{-1} = 0.01 \omega_{p}^{-1}/\beta $ in order to resolve the gyromotion. 
The range of run times indicated in the table are a result of the time it takes to reach convergence in the cumulative integral of the velocity autocorrelation functions, and were found to depend significantly on $\beta$. 

\begin{table}
\caption{Magnetized OCP Simulation Parameters}
\centering
\begin{tabular*}{0.475\textwidth}{c @{\extracolsep{\fill}}c c c}
\hline \hline
$\Gamma$ & $\beta$ & $t_{\text{run}}(\omega_p^{-1})$ \\
\hline
$0.1$ & $0.001-10$ & $3276.80-6553.60$ \\
$1$ & $0.01-3$ & $6553.60-8703.1808$ \\
$10$ & $0.001-4$ & $6553.60$ \\
$100$ & $0.01-10$ & $5242.88-6553.60$ \\
\hline \hline
\end{tabular*}
\label{table:Mag}
\end{table}

\section{Unmagnetized Plasma\label{sec:unmagnetized}} 

\begin{figure}
\includegraphics[width=8.5cm]{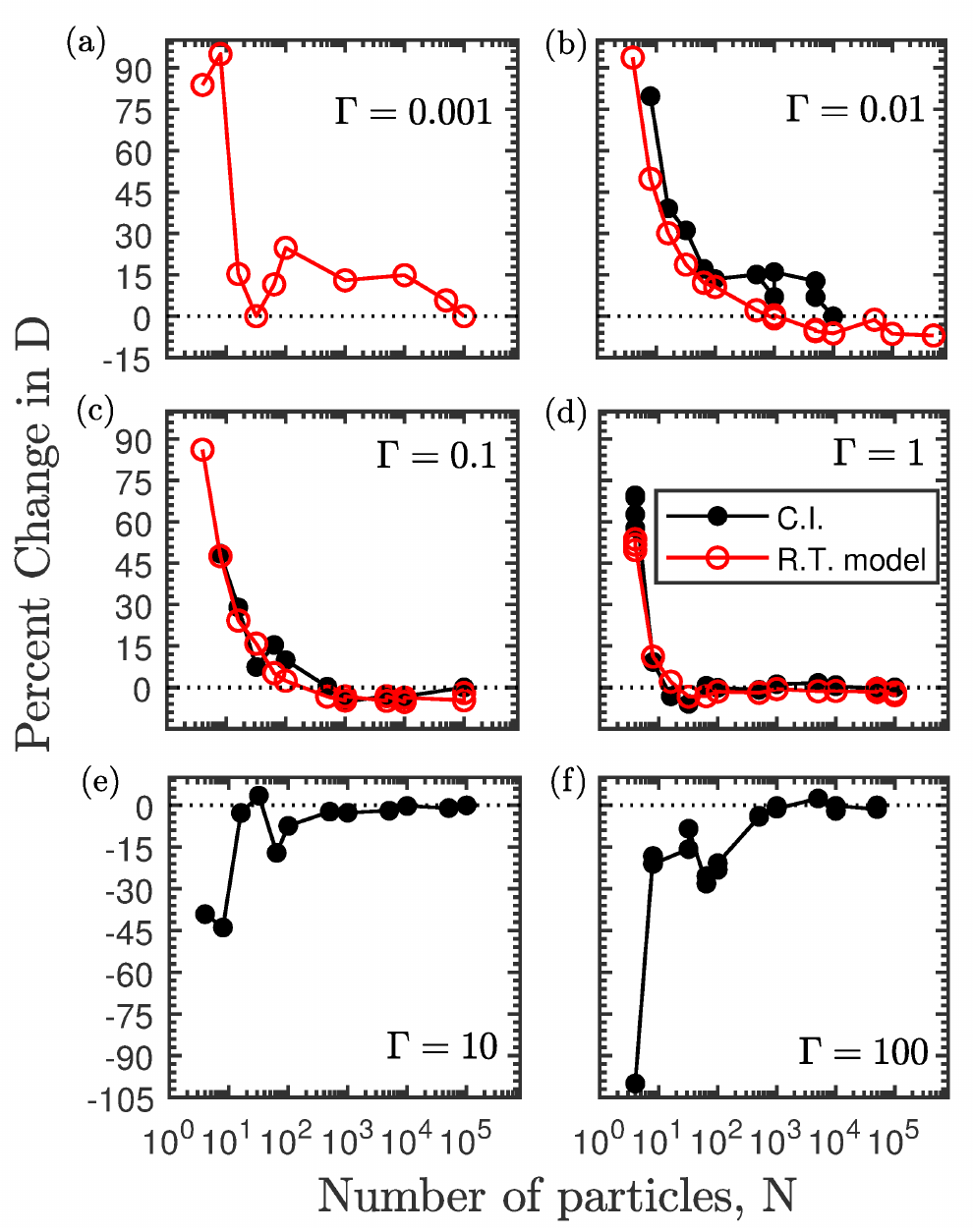}
\caption{Percent change of the computed diffusion coefficient versus the number of particles used in the simulation.  Percent change was calculated in reference to the result from the simulation that used the largest number of particles. For (a), this was from the relaxation time model (``R.T. model,'' red circles), and for (b)-(f) from the cumulative integrals (``C.I.'', black circles). }
\label{fig:PC_v_N}
\end{figure}

First, we review the correlation length and timescales for unmagnetized plasmas in order to set the expectations for comparison to strongly magnetized plasmas. 
Figure~\ref{fig:PC_v_N} shows a convergence test of the computed diffusion coefficients with respect to the simulation size, quantified in terms of the number of particles. 
Recall that since the density is fixed, the domain volume is proportional to the particle number, and the domain is cubic: $L^3 = N/n$. 

The expectation is that the simulation domain must significantly exceed the range over which particles interact.~\cite{Hockney:MD} 
At strongly correlated conditions ($\Gamma \gtrsim 1$), the interaction range is characterized by several interparticle distances, and increases gradually with $\Gamma$ as the dominance of plasmon oscillations become a predominant interaction mechanism.
This expectation is corroborated by the data, showing an increase in the number of particles required from just a few tens of particles at $\Gamma = 1$, to a few hundred particles at $\Gamma = 10$, and near one thousand particles at $\Gamma = 100$. 
At weakly correlated conditions ($\Gamma \lesssim 1$), the interaction range is characterized by the Debye length. 
Thus, the number of particles required is expected to scale steeply with decreasing coupling strength: $N \approx n \lambda_D^3 \propto \Gamma^{-3/2}$. 
This is consistent with the data shown in Fig.~\ref{fig:PC_v_N} where approximately $10^3$ particles were required for convergence at $\Gamma = 0.1$, but more than $10^4$ appear to be required to achieve better than $\sim 15\%$ accuracy at $\Gamma = 0.01$. 

\begin{figure}
\includegraphics[width=8.5cm]{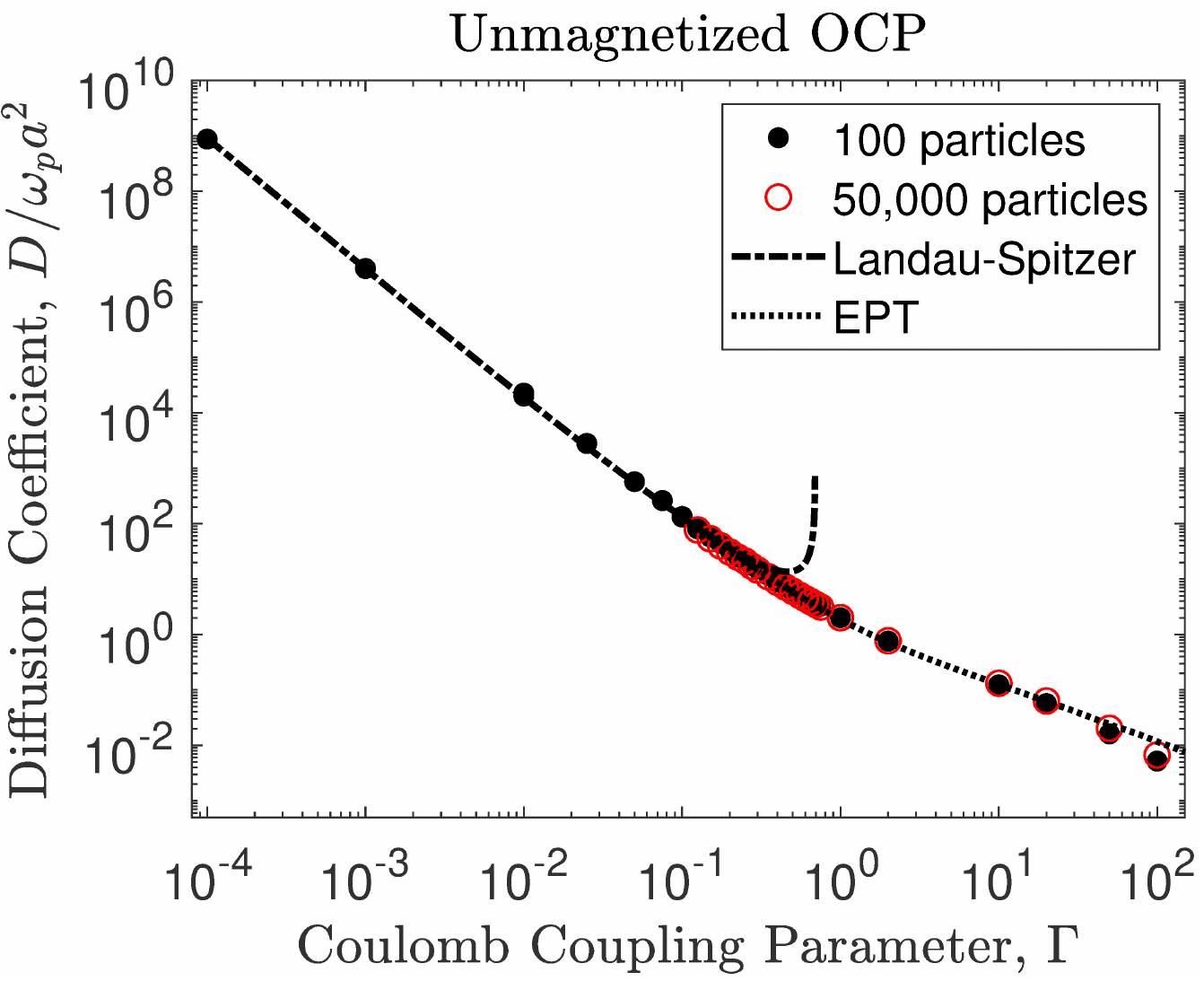}
\caption{Diffusion coefficients calculated from MD simulation across a range of coupling strengths, $\Gamma$.  Dots indicate MD simulations that used 100 particles, and red circles 50,000 particles.  The dash-dotted line represents predictions Landau-Spitzer theory,~\cite{Landau:1936, Spitzer:1953} and the dashed line Effective Potential Theory (EPT).~\cite{Baalrud:EPT}}
\label{fig:D_v_G}
\end{figure}

Another computational challenge at weak coupling is that many more timesteps are required. 
The timestep must be set to resolve close collisions: $\delta t \lesssim r_L/v_T \propto \Gamma^{3/2}/\omega_p$, where $r_L = e^2/k_\B T$ is the thermal distance of closest approach, or Landau length. 
Simultaneously, the Coulomb collision relaxation time increases so that the simulation time requirement also scales steeply with decreasing coupling strength: $\tau \propto \Gamma^{-3/2} \omega_p$. 
Putting these together, the number of timesteps required scales as $\tau/\delta t \sim \Gamma^{-3}$. 
The combination of the large particle number requirement and the long simulation duration requirement lead to the conclusion that MD simulations quickly become infeasible for $\Gamma \ll 1$. 
Our simulations were not able to reach convergence based on the cumulative integral method for $\Gamma \lesssim 0.01$. 

\begin{figure*}
\includegraphics[width=17cm]{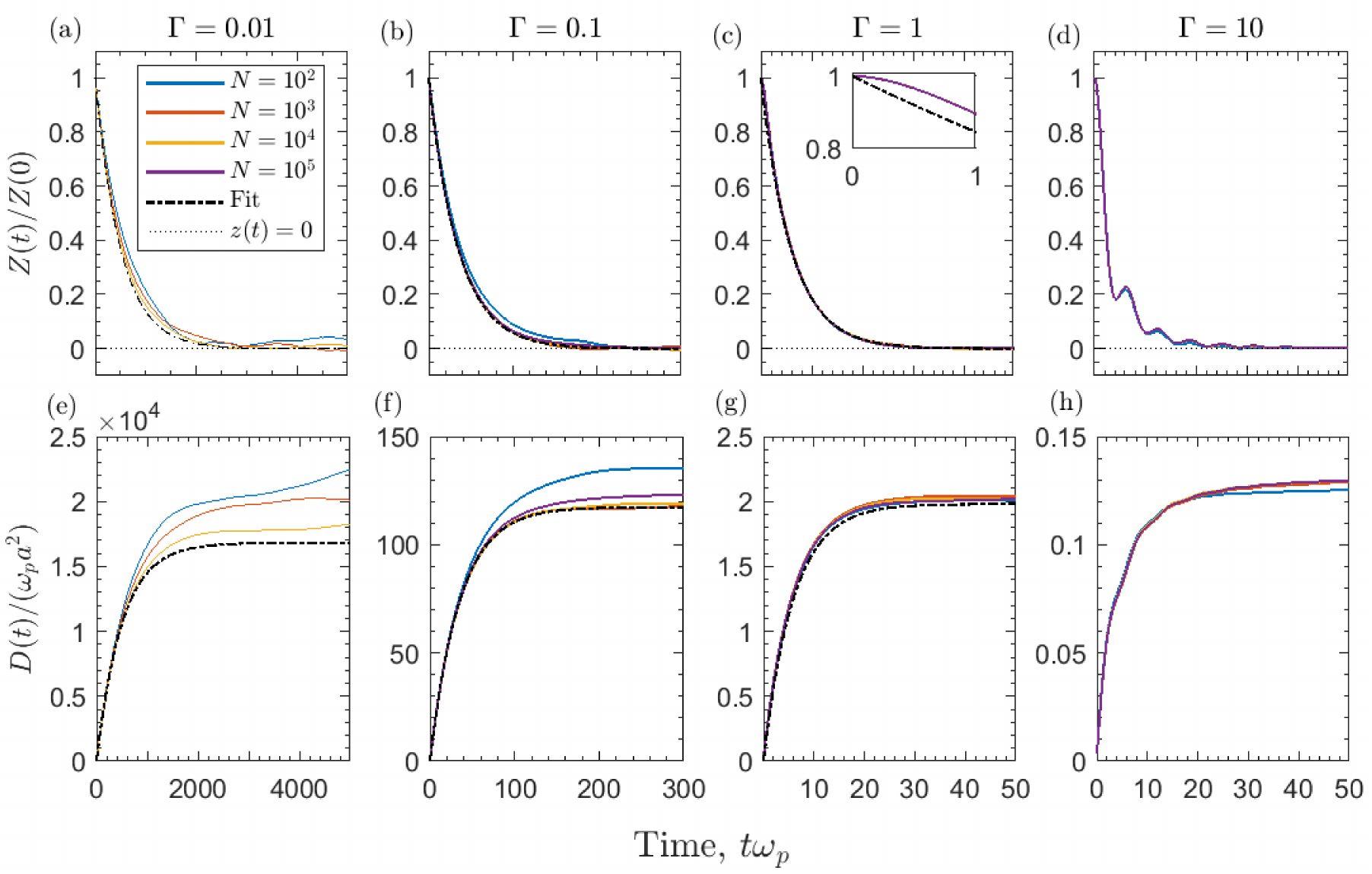}
\caption{The normalized velocity autocorrelation function, $Z(t)/Z(0)$, and the cumulative integral, $D(t)$, obtained from MD simulations with varied number of particles, $N$, for $\Gamma=0.01$ ((a) and (e)), $\Gamma=0.1$ ((b) and (f)), $\Gamma=1$ ((c) and (g)), and $\Gamma=10$ ((d) and (h)).  The dash-dotted line is the result of the fit from equation (\ref{Zt:eq}) to the $Z(t)$ data and is only present in the $\Gamma=0.01$, $\Gamma=0.1$, and $\Gamma=1$ plots.}
\label{fig:ZtandDt}
\end{figure*}

Despite these strict limitations for first-principles simulations, shortcuts can be made by utilizing known properties of weakly coupled plasmas. 
First, the simulation domain size is only weakly dependent on the particle number at weak coupling. 
Figure~\ref{fig:PC_v_N} shows that although the number of particles required for strict convergence to the few percent level is extremely large at weak coupling, simulations with only 100 particles were within $\sim 20\%$ of the expected result from standard theory. 
Figure~\ref{fig:D_v_G} shows a surprising result that this trend extends to very weak coupling, as convergence to the tens of percent level were obtained with only 100 particles down to $\Gamma = 10^{-4}$. 
Here, the data is compared with the accepted theoretical predictions resulting from the lowest order Chapman-Enskog solution to the Boltzmann equation~\cite{Chapman:1991}
\begin{equation} \label{D_CE:eq}
\frac{D}{\omega_p a^2} = \frac{\sqrt{\pi/3}}{\Gamma^{5/2} \Xi},
\end{equation}
where $\Xi \rightarrow \ln{\Lambda}=\ln{ ( \lambda_D/r_{L} )}$ corresponds to the traditional Landau-Spitzer formula applicable to  the weakly coupled limit.~\cite{Landau:1936, Spitzer:1953}  
Here, the Effective Potential Theory (EPT) has been applied in order to extend the traditional theory to higher coupling ($\Gamma \lesssim 20$) by calculating $\Xi$ using the potential of mean-force,~\cite{Baalrud:EPT} but the predictions are equivalent for $\Gamma \lesssim 0.1$. 

The reason for the relatively accurate results at low particle number is likely associated with the Ewald summation technique. 
In this method, long-range interactions are computed as the average charge interpolated to a mesh, which extends beyond the nominal simulation domain size by making use of periodically replicated cells. 
Since long range interactions are weak, and are weakly correlated with the particle for which the force is being computed, the charge on the grid is an accurate representation of the long-range interaction. 

Further reductions in computational effort can be made by utilizing known properties for the time dependence of the relaxation of the velocity autocorrelation function. 
Namely, the weak interactions and Markovian behavior justify the relaxation time approximation in which the equation of motion for particles can be approximated by a Langevin form~\cite{Hansen:2013}
\begin{equation} \label{langevin:eq}
m \frac{d \vc{v}(t)}{d t} = -m \xi \vc{v}(t) + \vc{R}(t)
\end{equation}
with a constant friction coefficient, $\xi$, and an assumption that successive collisions are uncorrelated, $\langle \vc{R}(t) \cdot \vc{v}(0) \rangle = 0$.
With these assumptions, Eq.~(\ref{eq:Z}) reduces to a simple exponential form 
\begin{equation} \label{Zt:eq}
Z(t) = Z(0) \exp{(-t \xi)},
\end{equation}
and the resulting self-diffusion coefficient is
\begin{equation} \label{D:eq}
\frac{D}{\omega_p a^2} = \frac{\omega_p}{3 \Gamma \xi}.
\end{equation}
Figure~\ref{fig:ZtandDt} shows that the simulated velocity autocorrelation functions can be well fit by Eq.~(\ref{Zt:eq}) with $\xi$ as the fit parameter when $\Gamma \lesssim 1$, and Fig.~\ref{fig:D_v_G} shows that the diffusion coefficients resulting from using the fit $\xi$ value in Eq.~(\ref{D:eq}) agree well with the theoretical predictions.
Figure~\ref{fig:PC_v_N} shows this relaxation time approximation model agrees well with diffusion coefficients obtained from the cumulative integral when $\Gamma \lesssim 1$.

\begin{table*}
\caption{Summary of transport regimes.~\cite{Baalrud:regimes} Here, $\lambda_{\text{col}} =$ denotes the collision mean-free-path, $a=$ the average interparticle spacing, and $r_{L}=$ the thermal distance of closest approach (Landau length).}
\centering
\begin{tabular*}{1\textwidth}{c @{\extracolsep{\fill}}c c c}
\hline \hline
Region & Degree of Magnetization & Gryoradius Criteria & $\beta-\Gamma$ Criteria \\
\hline
$1$ & Unmagnetized & $r_c \gtrsim \lambda_{\text{col}}$ & $\beta \lesssim 0.32 \Gamma^{3/2} \Xi(\Gamma)$ \\
$2$ & Weakly Magnetized & max $\{ \lambda_D$,$a/ \sqrt{2}\} \lesssim r_c \lesssim \lambda_{\text{col}}$ & $0.32 \Gamma^{3/2} \Xi(\Gamma) \lesssim \beta \lesssim \text{min} \{ 1, \sqrt{2/(3 \Gamma)} \}$ \\
$3$ & Strongly Magnetized &  $r_L \lesssim r_c \lesssim$ max $\{ \lambda_D$,$a/ \sqrt{2}\}$ & $1 \lesssim \beta \lesssim 1/\sqrt{6 \Gamma^3}$ \\
$4$ & Extremely Magnetized & $r_c \lesssim \text{min} \{r_L, a/\sqrt{2}, \lambda_{\text{col}} \}$ & $\text{max} \{1/\sqrt{6 \Gamma^3}, \sqrt{2/(3 \Gamma)}, 0.32 \Gamma^{3/2} \Xi \} \lesssim \beta$ \\
\hline \hline
\end{tabular*}
\label{table:Regimes}
\end{table*}

Making use of the exponential decay approximation allows a significant decrease of the total computation time, because the simulations need only be run long enough to capture the slope of the decay. 
This typically requires total simulation lengths that are several plasma periods, as the first plasma period consists of non-Markovian behavior associated with initial ballistic motion, as shown in the inset plot of Fig.~\ref{fig:ZtandDt}. 

Although these shortcuts make MD simulations more tractable at weak coupling, it should be noted that this is only possible because the underlying physics processes are known. 
Since MD simulations are most often utilized as a first-principles approach in cases that the physical processes are not well understood, the simplifications are of limited usefulness. 

\section{Magnetized Plasmas\label{sec:magnetized}} 

Previous MD simulations of the magnetized OCP have identified four regimes in which different underlying physical processes lead to qualitative changes in the diffusion coefficients.~\cite{Baalrud:regimes} 
These are summarized in Table~\ref{table:Regimes}, and can be understood based on a comparison of the gyroradius to other length scales relevant to collision physics: 
(1) Unmagnetized regime: When the gyroradius is larger than the collision mean free path, particles do not gyrate before scattering, so the magnetic field does not significantly influence collisional transport. 
(2) Weakly magnetized regime: Here, the magnetic field influences transport because the gyroradius is smaller than the collision mean free path, but it does not influence collisions at the microscopic scale because the gyroradius is larger than the Debye length.  This is the traditional regime of magnetized weakly coupled plasmas, as described by Braginskii transport theory.~\cite{Braginskii:1965} 
(3) Strongly magnetized regime: Here, interactions between particles at the microscopic scale are influenced by the magnetic field, since the gyroradius is smaller than the Debye length, but the gyroradius is much larger than the distance of closest approach in a binary collision. 
Boltzmann kinetic theory does not apply in this regime, and generalized collision models must be used.~\cite{DubinPOP2014,Jose:2020} 
(4) Extremely magnetized regime: When the gyromotion is smaller than any other scale relevant to collision physics, particles move almost in one-dimension, but undergo nearly 180$^\circ$ degree collisions at the ends of collision cylinders. 

Previous work has explored the diffusion coefficient in each of these regimes.~\cite{Ott:2011,Baalrud:regimes}  
Here, we focus primarily on the correlation scales associated with particle interactions in regions (3) and (4). 

\subsection{Correlation length\label{sec:length}} 

Strong magnetization causes a channeling effect in which particles move a long distance parallel to the magnetic field within channels with a width characterized by the gyroradius.
This causes a long-range spatial correlation that can be observed by studying the convergence of velocity autocorrelation functions and self-diffusion coefficients with respect to the simulation domain size. 

\begin{figure*}
\includegraphics[width=15cm]{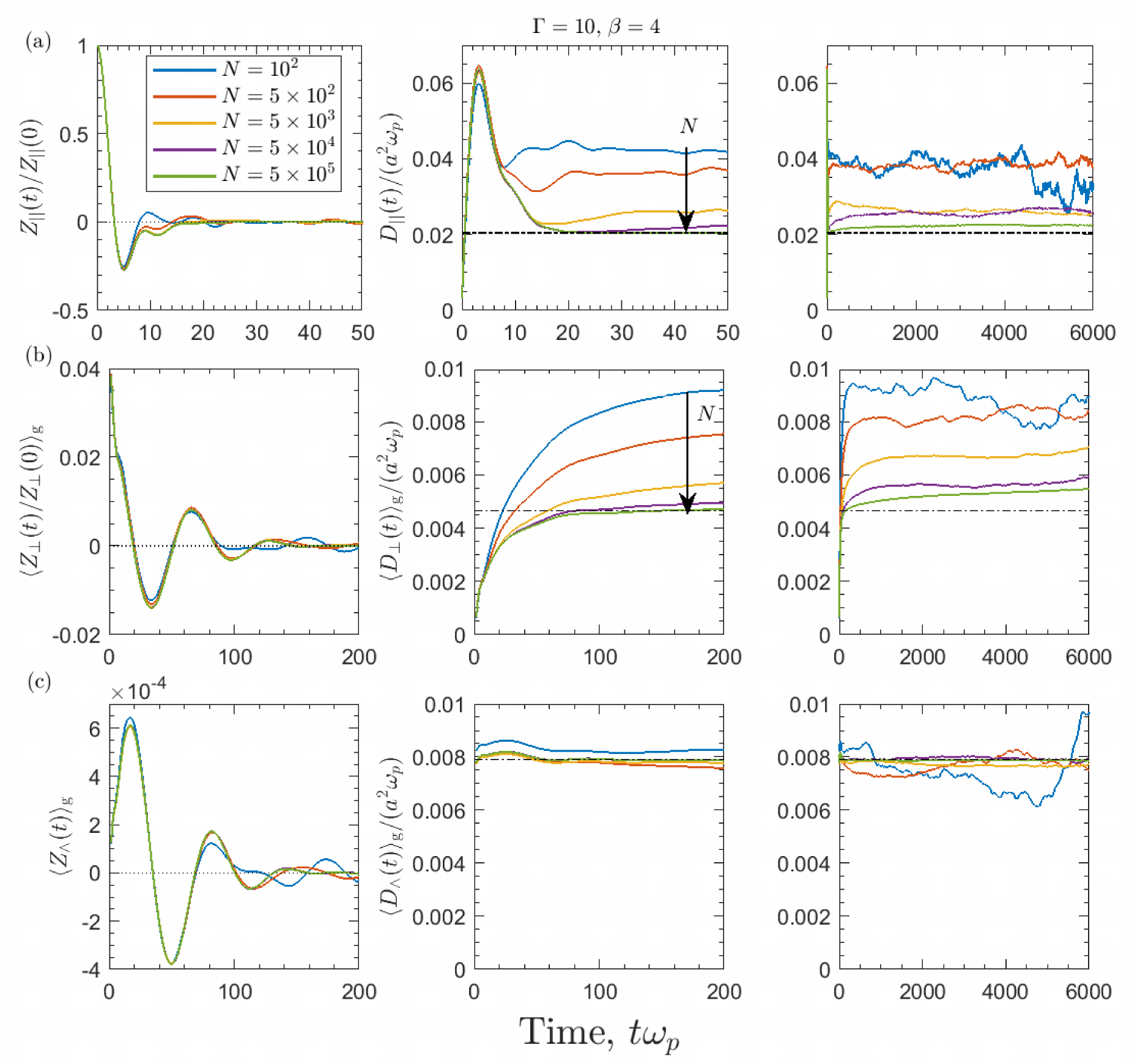}
\caption{The velocity correlation functions ($Z_{\parallel}(t)/Z_{\parallel}(0)$, $\langle Z_{\bot}(t)/Z_{\bot}(0) \rangle_{\text{g}}$, and $\langle Z_{\wedge}(t) \rangle_{\text{g}}$) and cumulative integral ($D_{\parallel}(t)$, $\langle D_{\bot}(t) \rangle_{\text{g}}$, and $\langle D_{\wedge}(t) \rangle_{\text{g}}$) at $\Gamma=10$ and $\beta=4$ for the (a) parallel, (b) perpendicular, and (c) transverse directions.  The right panel plots show the cumulative integrals at an extended time scale.  The $\langle ... \rangle_{\text{g}}$ indicates the quantity was gyroaveraged.  The dash-dotted line is a horizontal reference line.  The arrow indicates the increasing $N$ trend of the plots. }
\label{fig:G10_b4}
\end{figure*}

Figure~\ref{fig:G10_b4} shows computations of the velocity autocorrelation functions for $\Gamma = 10$ and $\beta = 4$ (region 4) as the number of particles in the simulation increases from 100 to 500,000. 
Here, $Z_\perp$ and $Z_\wedge$ have been gyroaveraged in order to visualize the behavior of the decay, since each of these components has a rapid oscillation at the gyrofrequency that otherwise obscures the general trends; see Ref.~\onlinecite{Baalrud:regimes}. 
The figure shows that a very large particle number is required for convergence in both the parallel and perpendicular directions, exceeding $10^5$ particles. 
This is much larger than the approximately 500 particles required to reach convergence in the unmagnetized regime at the same $\Gamma$ value; see Fig.~\ref{fig:PC_v_N}. 
This slow convergence is also clear in the cumulative integrals shown in the middle column. 
The self-diffusion coefficients are computed from the plateau of these curves, demonstrating that the errors in the velocity autocorrelation functions add so that the error associated with insufficient particle number is stark. 
This error is exacerbated by the additional observation that the true plateau associated with diffusion is delayed until very late times, as shown in the right column of the figure. 
Section~\ref{sec:time} discusses how this delay in the approach to a diffusive timescale is related to the extended spatial correlation. 

\begin{figure*}
\includegraphics[width=17cm]{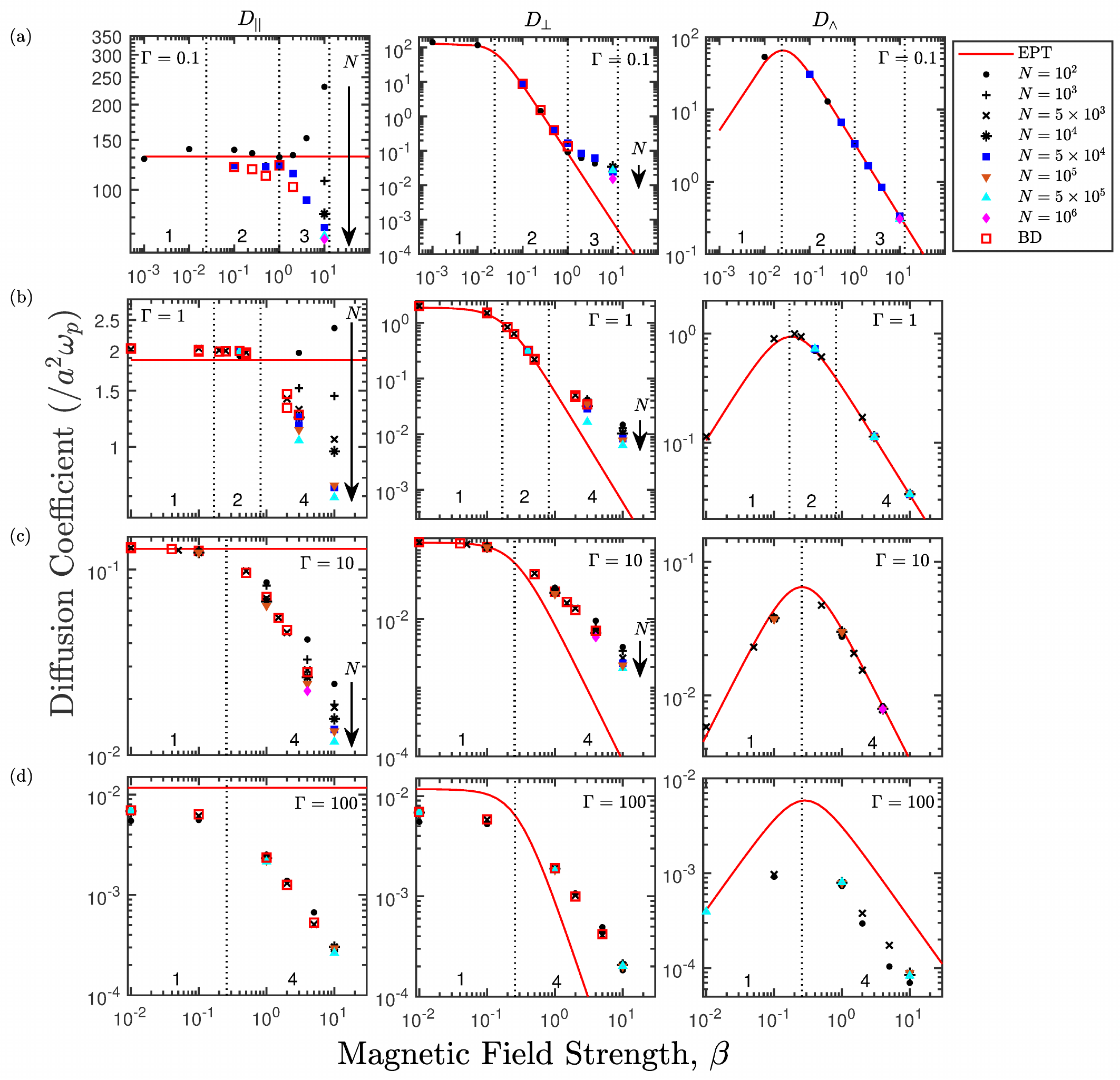}
\caption{Parallel ($D_{\parallel}$), perpendicular ($D_{\bot}$), and transverse ($D_{\wedge}$) diffusion coefficients for (a) $\Gamma=0.1$, (b) $\Gamma=1$, (c) $\Gamma=10$, and (d) $\Gamma=100$.  The red solid line shows the prediction of the EPT theory, and the open red squares, labeled BD, are results from Ref.~\onlinecite{Baalrud:regimes}.}
\label{fig:all_Dvb}
\end{figure*}

Figure~\ref{fig:all_Dvb} shows the self-diffusion coefficients computed from the late-time plateau in the cumulative integral of the velocity autocorrelation function over a broad range of Coulomb coupling strengths ($\Gamma = 0.1, 1, 10$ and 100) and magnetization strengths ($\beta = 10^{-3} - 10$). 
This range of conditions spans all four of the magnetization regimes identified in table~\ref{table:Regimes}, which are indicated as the regions between the vertical dotted lines in the figure. 
These results confirm the expectation from Sec.~\ref{sec:unmagnetized} that only a few hundred particles are required to obtain convergence in regions 1 and 2. 
However, as the magnetic field strength increases into regions 3 or 4, the $D_\parallel$ and $D_\perp$ coefficients depend significantly on the number of particles used and more particles are required as $\beta$ increases. 
This trend is most pronounced in $D_\parallel$ at weak coupling, but it is also significant in the $D_\perp$ component. 
For example, over $10^6$ particles are required to reach convergence at $\Gamma = 0.1$ and $\beta = 10$. 
The spread of computed coefficients over the range of $N$ values narrows as $\Gamma$ increases, becoming insensitive to the number of particles used over this range of $\beta$ when $\Gamma = 100$. 
The $D_\wedge$ component is primarily associated with the Hall effect, which is insensitive to interactions, and this is likely why this component is also insensitive to the number of particles used. 

The figure also shows a comparison with theory (red line). 
The unmagnetized and weakly magnetized regimes (1 and 2) are expected to be well-described by the traditional Braginskii transport theory~\cite{Braginskii:1965} when $\Gamma \ll 1$.  
The limitation to small coupling strengths is because the plasma kinetic theory upon which this solution is derived applies only to this regime.  
However, the recent effective potential theory (EPT) has provided a method to extend plasma kinetic theory to the regime of $\Gamma \lesssim 20$.~\cite{Baalrud:EPT}   
To lowest order in the Chapman-Enskog expansion, the diffusion coefficients predicted by this model are  
\begin{subequations}
\begin{eqnarray}
\label{D_paraEPT:eq}
\frac{D_{\parallel}}{\omega_p a^2} &=& \frac{\sqrt{\pi/3}}{\Gamma^{5/2} \Xi} \\ 
\label{D_perpEPT:eq}
D_{\bot} &=& \D_{\parallel} \bigg( 1 + 3 \pi \frac{\beta^2}{\Gamma^3 \Xi^2} \bigg)^{-1} \\ 
 \label{D_wedgeEPT:eq}
D_{\wedge} &=& \frac{\sqrt{3 \pi} \beta}{\Gamma^{3/2} \Xi} D_{\bot}
\end{eqnarray}
\end{subequations}
where the generalized Coulomb logarithm, $\Xi$, is calculated from the potential of mean force as described in Ref.~\onlinecite{Baalrud:EPT}.  
The results of this model are shown as the red line in Fig.~\ref{fig:all_Dvb}. 

As expected, good agreement is observed in the regimes in which the theory is expected to apply: regions 1 and 2, and for $\Gamma \lesssim 20$. 
For stronger magnetization strengths, in regions 3 and 4, the MD results diverge from these predictions, with the $D_\parallel$ coefficient becoming dependent on $\beta$ and the $D_\perp$ coefficient scaling more gradually with $\beta$ than the $\beta^{-2}$ dependence predicted by the theory. 
These observations regarding $D_\parallel$ and $D_\perp$ have previously been made in Refs.~\onlinecite{Ott:2011} and \onlinecite{Baalrud:regimes}. 
Figure~\ref{fig:all_Dvb} shows the first MD computations of $D_\wedge$ over this range of parameters. 
The good agreement with theory even in regions 3 and 4 is likely because the $D_\wedge$ coefficient is determined entirely by the Hall effect in this region, and becomes independent of the collision model. 

\begin{figure}
\includegraphics[width=8.5cm]{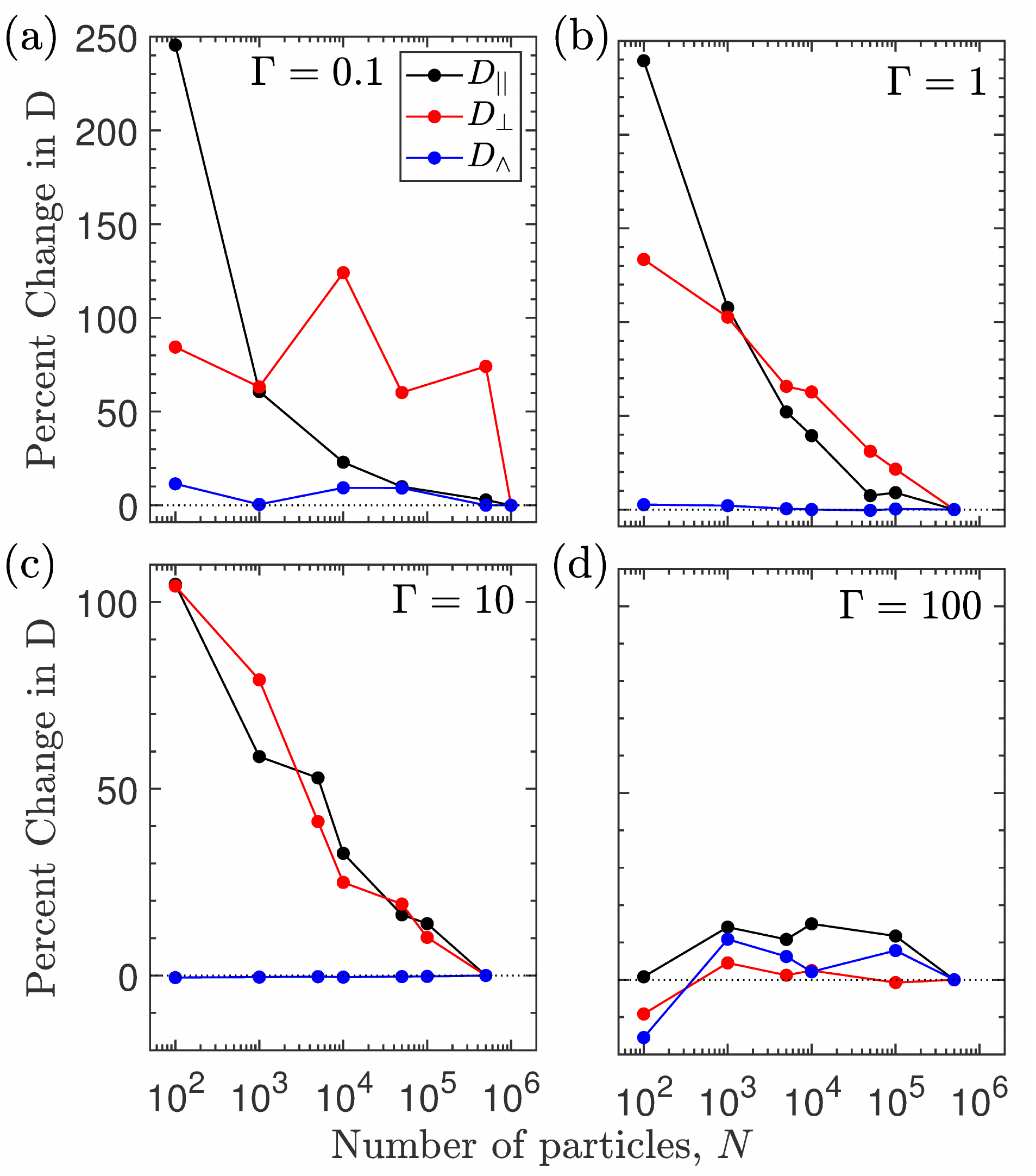}
\caption{Percent change in diffusion coefficients in reference to the values calculated from the simulation with the largest number of particles, $N$, for (a) $\Gamma=0.1$, (b) $\Gamma=1$, (c) $\Gamma=10$, and (d) $\Gamma=100$ at $\beta=10$.}
\label{fig:DvN}
\end{figure}

Next, we return to the question of why the MD simulations require a very large domain to reach convergence when the plasma is strongly magnetized, and weak to moderately coupled. 
Figure~\ref{fig:DvN} shows this trend from a different perspective, plotting the self-diffusion coefficients referenced to the value computed from the simulation with the largest particle number. 
This shows even more clearly than Fig.~\ref{fig:all_Dvb} that an extremely large number of particles is required at strong magnetization when $\Gamma$ is moderate to small ($\Gamma \lesssim 10$). 
Plotted this way, the data demonstrates that convergence to within a few percent would take many more than $10^6$ particles; extrapolating the observed trends suggests that it may take more than $10^8$. 

\begin{figure}
\includegraphics[width=8.5cm]{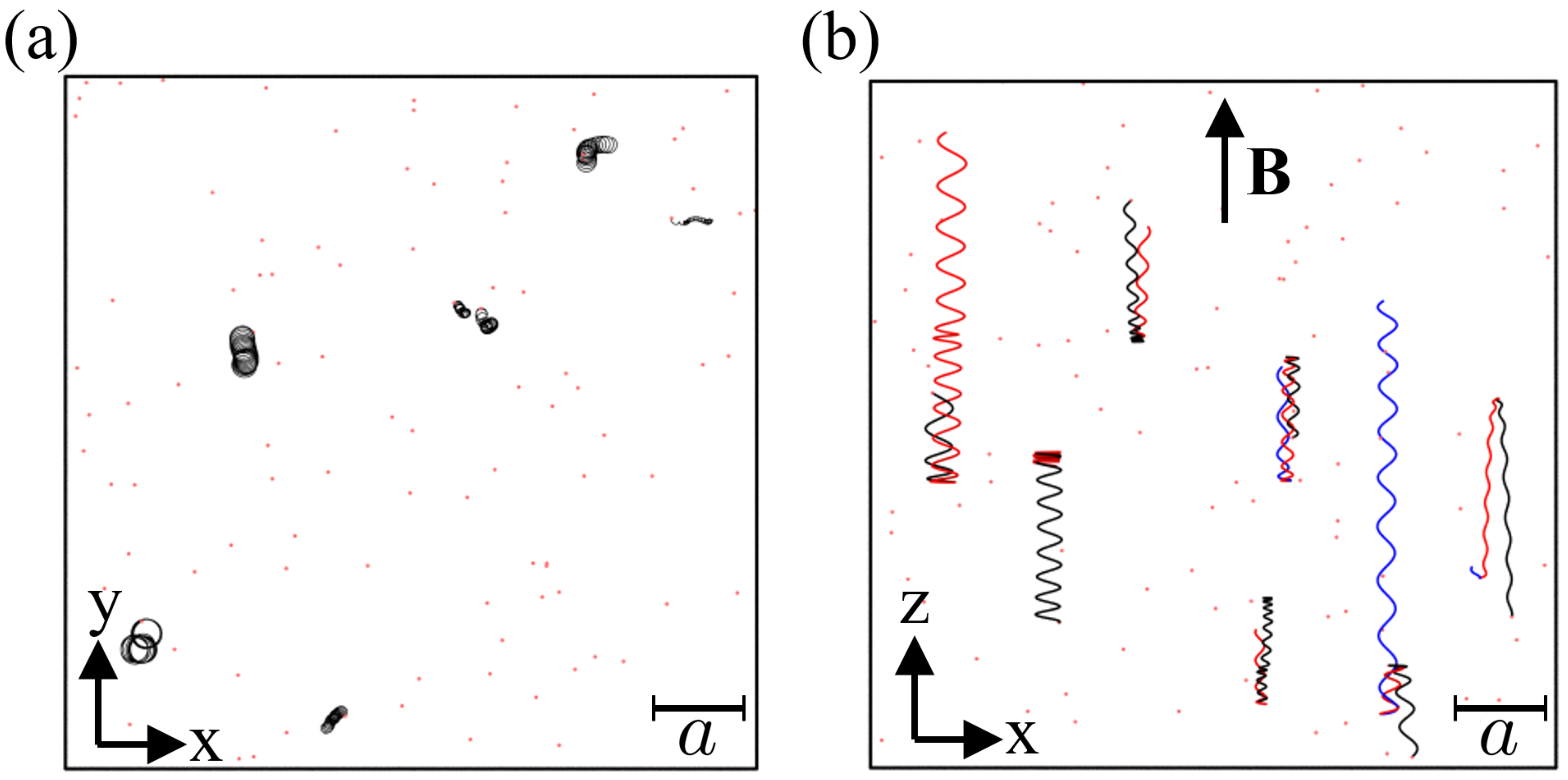}
\includegraphics[width=8.5cm]{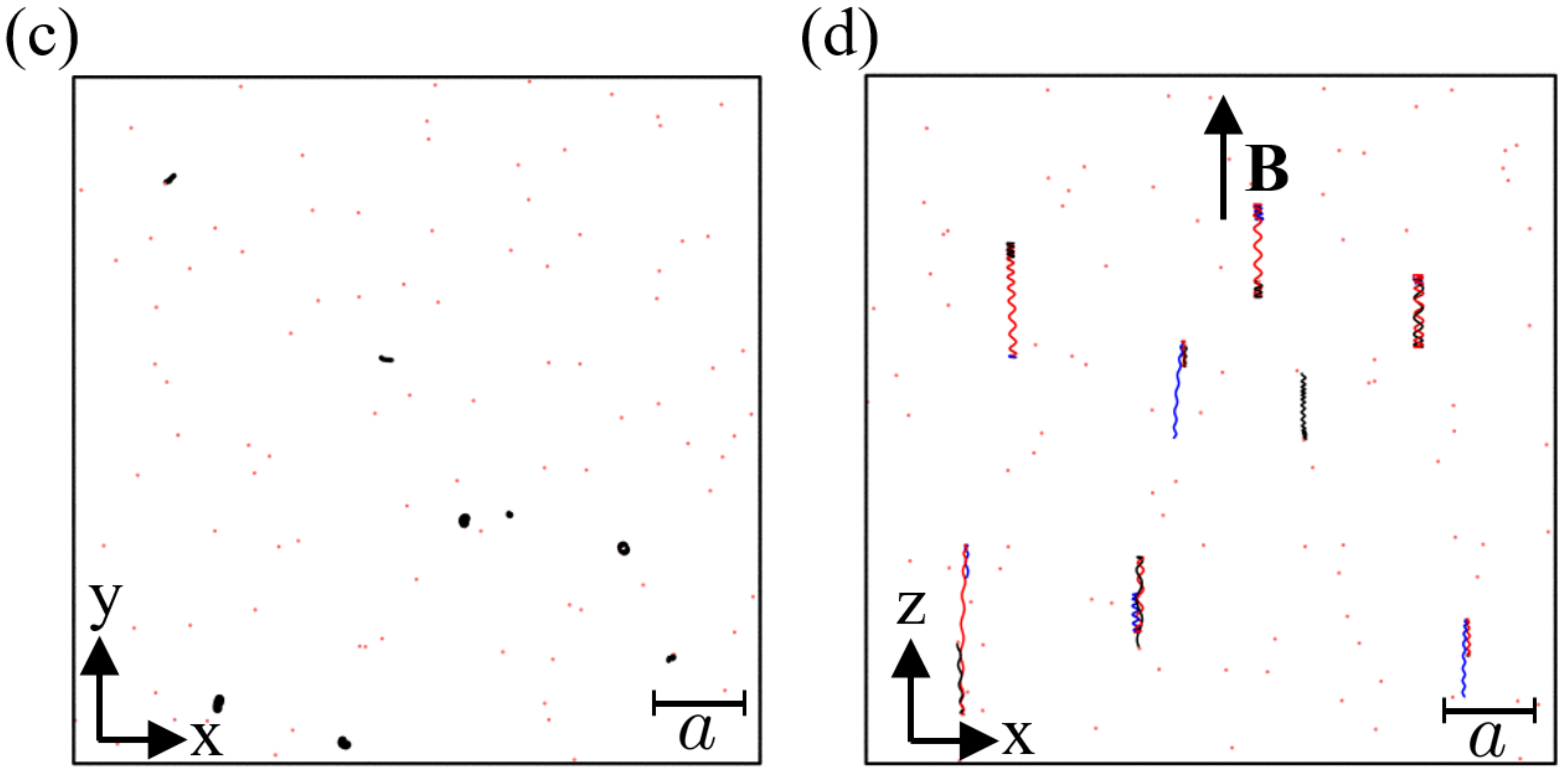}
\includegraphics[width=8.5cm]{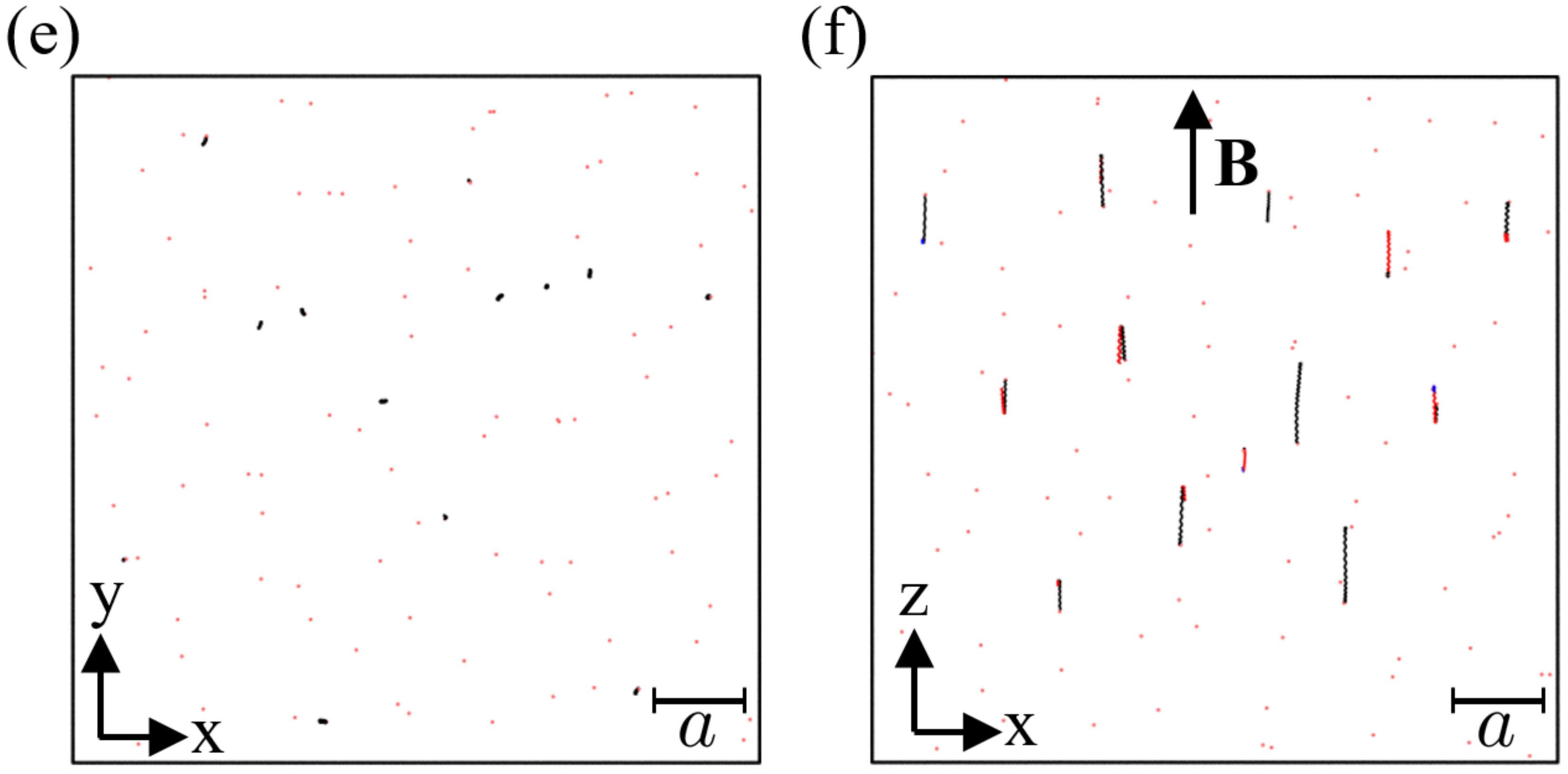}
\caption{Trajectories of a select few particles at $\beta=10$, $t_{\text{run}}=10 \omega^{-1}_{p}$, and $N=100$ particles with $\vc{B} = B \hat{\vc{z}}$ for (a)-(b) $\Gamma=1$, (c)-(d) $\Gamma=10$, and (e)-(f) $\Gamma=100$.  The trajectories in the $z$-$x$ plane ((b),(d), and (f)) change color when the particle underwent a $180^\circ$ collision.  The dimension of the simulation box was $L=7.5a$.  (This figure was created using OVITO~\cite{ovito}).}
\label{fig:trajectories}
\end{figure}

The physical mechanism responsible for this long-range correlation is revealed by looking at the particle trajectories. 
Figure~\ref{fig:trajectories} shows example trajectories from a simulation at strongly magnetized conditions: $\Gamma = 1$ and $\beta=10$. 
The most apparent feature is that particles move a long distance in the direction along the magnetic field, before scattering with a neighboring particle on a nearby field line, at which point it undergoes a 180$^\circ$ reflection. 
In this way, particles are largely confined to a ``collision cylinder'' with a length of approximately the mean free path for a 180$^\circ$ scattering event and a width of approximately a gyroradius. 
However, particles are also observed to continuously and gradually drift across the magnetic field. 
This aspect of the motion is associated with the continuous weak interaction between charged particles through the long-range Coulomb force. 
The result is that the ``collision cylinders'' broaden, and even migrate, in time. 
Thus, both large-angle reflective scattering associated with close interactions, and the small-angle continuous drifting associated with long-range interactions contribute to the overall transport. 
As the coupling strength increases, the plasma becomes more collisional and the distance between 180$^\circ$ scattering events shrinks considerably. 
In the strongly coupled regime, particles are continuously undergoing strong inter-particle interactions and these prevent particles from migrating far along the magnetic field, even in a strong magnetic field regime. 
This reduces the spatial correlation length in the parallel dimension, and a correspondingly smaller simulation domain is required to reach convergence.

In the strongly magnetized regime, the most stringent requirement for convergence appears to be that the parallel dimension of the simulation domain be much larger than the parallel correlation scale separating 180$^\circ$ scattering events. 
Since the boundaries are periodic, part of this requirement is that the same two particles do not interact as a consequence of traveling through the domain boundary, as this would cause an artificial correlation. 
However, determining the domain size requirement is not as simple as estimating the intersection of a collision cylinder of gyroradius width intersecting nearest neighbor particles because the cross-field drift due to long-range interactions causes collision cylinders to migrate. 

\begin{figure}
\includegraphics[width=8.5cm]{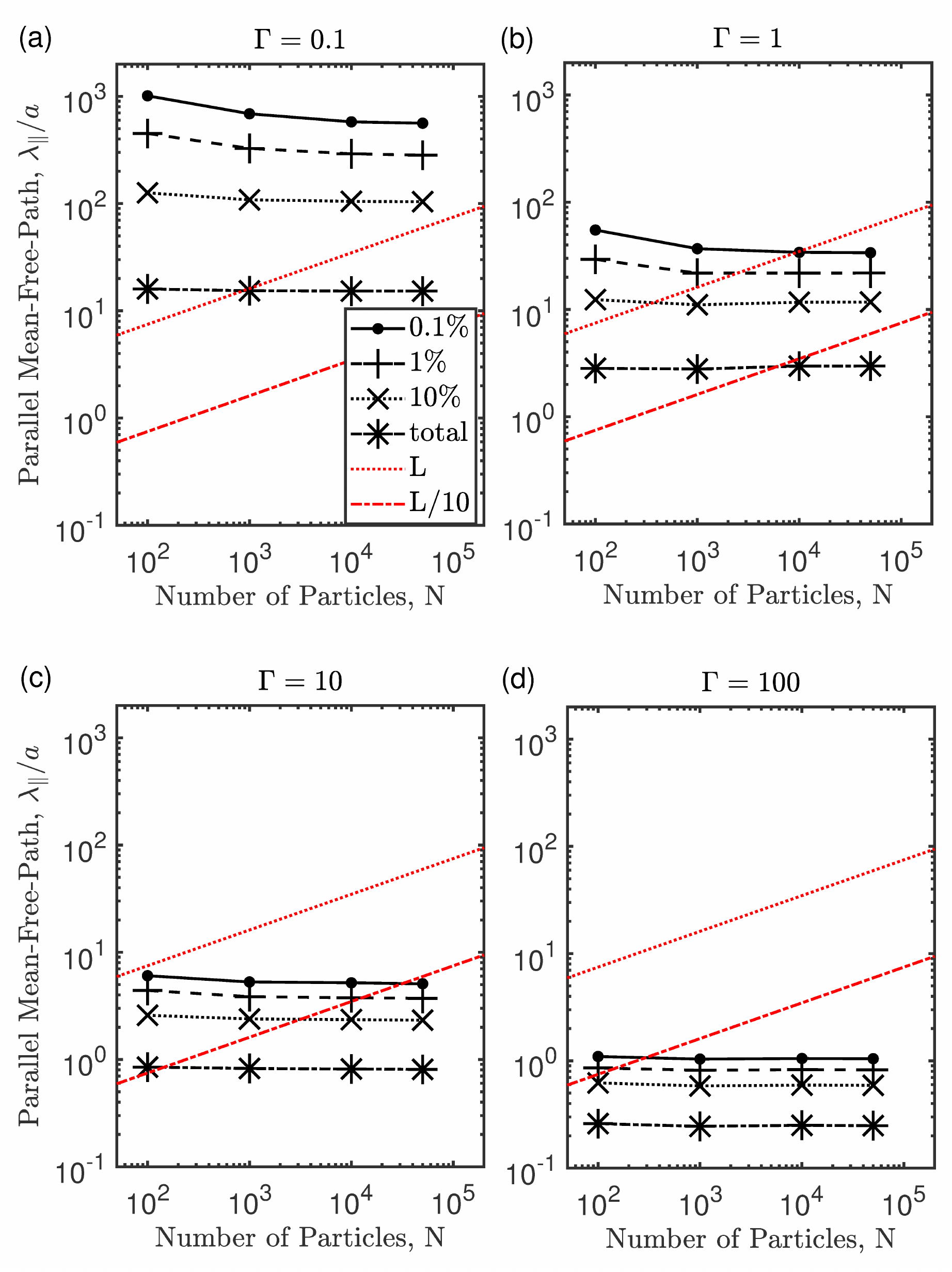}
\caption{The parallel mean-free-path for (a) $\Gamma=0.1$, (b) $\Gamma=1$, (c) $\Gamma=10$, (d) $\Gamma=100$.  All simulations were at a $\beta=10$ with $t_{\text{run}}=1000 \omega^{-1}_p$.  The red dotted line shows the dimension of the the simulation box, $L$, and the red dashed line shows $L/10$. }
\label{fig:mfp}
\end{figure}

A test of this hypothesis is provided in Fig.~\ref{fig:mfp}. 
This shows the mean free path for 180$^\circ$ scattering events along the magnetic field for $\beta = 10$ and coupling strengths $\Gamma = 0.1, 1, 10$ and 100 as a function of the simulation domain size.  
Data labeled ``total'' included all particles in the mean free path calculation, whereas those labeled $0.1\%$, $1\%$, and $10\%$ included only the longest $0.1\%$, $1\%$, and $10\%$ mean free-paths, respectively.
The figure shows that the mean free path, in units of average interparticle distance, decreases as $\Gamma$ in increases, as is expected from the higher collisionality at strong coupling. 
Red lines in the figure denote the length of the domain (dotted line), and $1/10$ of the domain (dashed line). 
The notion that the vast majority of collision mean free paths ($\lesssim 0.1$) have to be much smaller than the simulation domain ($L/10$) leads to an expectation that is consistent with that observed in the convergence tests of Fig.~\ref{fig:DvN}. 

These results indicate that in a strongly magnetized plasma, the correlation strength depends not only on the Coulomb coupling parameter ($\Gamma$), but also on the magnetization parameter ($\beta$). 
Indeed, it has been pointed out previously that strong magnetization can cause both positive and negative correlations that are indicative of a $\Gamma > 1$ regime, even when $\Gamma < 1$.~\cite{Baalrud:regimes} 
Although the correlation scale of these interactions is long, they are also associated with large-angle (180$^\circ$) scattering events, unlike the weak long-range interactions observed for $\Gamma \ll 1$. 
For this reason, the Ewald summation technique does not alleviate the requirement on the size of the simulation domain as it does in a weakly coupled plasma; see Sec.~\ref{sec:unmagnetized}. 
The domain must be large enough to resolve particle-particle interactions over the extended parallel correlation scale.

\begin{figure*}
\includegraphics[width=17cm]{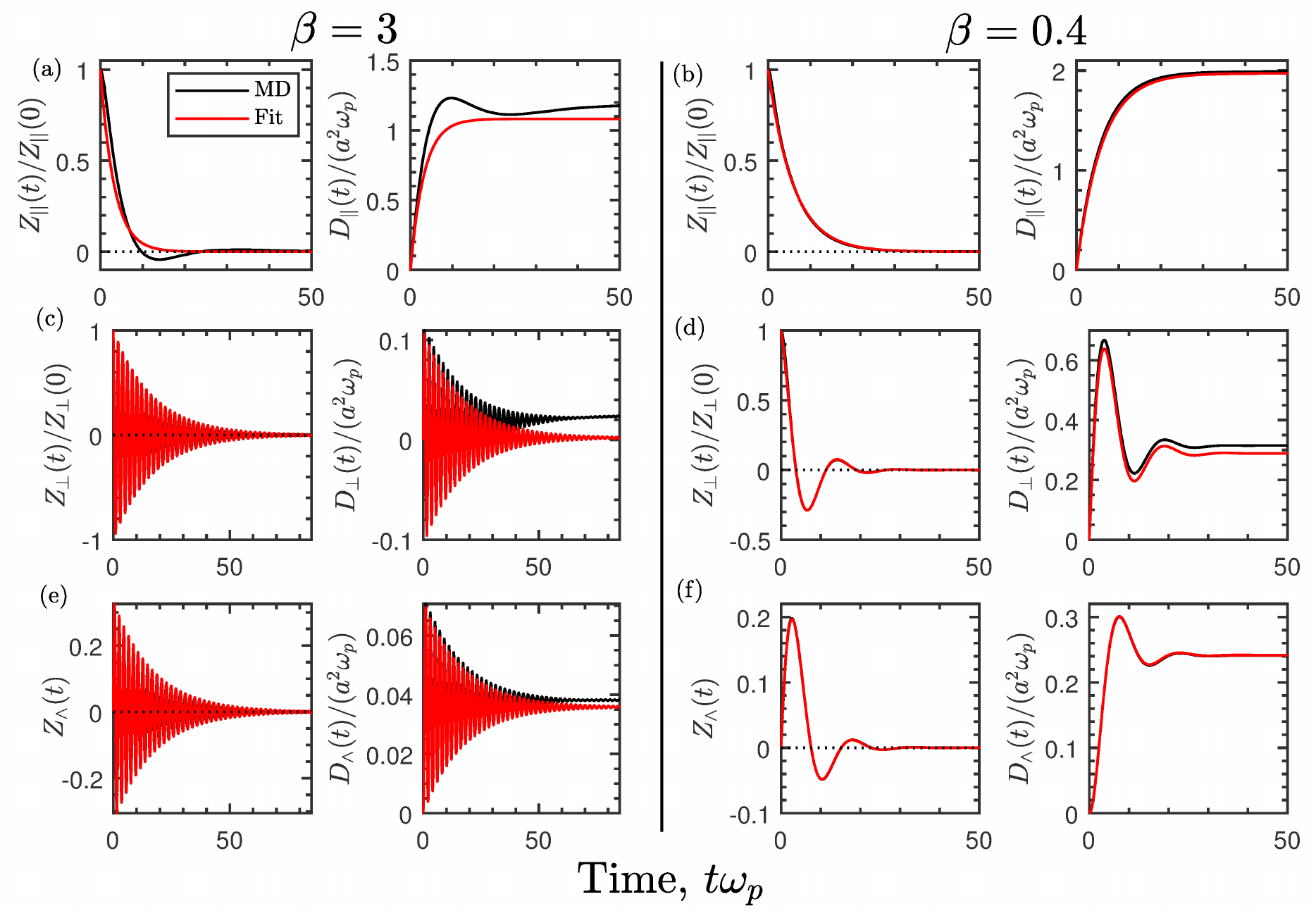}
\caption{The parallel [(a) and (d)], perpendicular [(b) and (e)], and transverse [(c) and (f)] velocity correlation functions, $Z(t)$, and cumulative integrals, $D(t)$, at $\Gamma=1$, $\beta=3$, and $N=5 \times 10^{3}$.  The red line is the result of the fit to the MD data using Eqs. (\ref{Zpara:eq}).}
\label{fig:mag_g1_VACF}
\end{figure*} 

Similarly, the importance of this magnetization-induced correlation renders the relaxation time approximation invalid, even at small $\Gamma$. 
The magnetized plasma version of Eq.~(\ref{langevin:eq}) includes the Lorentz force
\begin{equation} \label{magLangevin:eq}
m \frac{d \vc{v}(t)}{d t} = -m \underline{\underline{\xi}} \cdot \vc{v}(t) + q(\vc{v} \times \vc{B}) + \vc{R}(t)
\end{equation}
where the constant friction coefficients are now represented as a tensor
\begin{equation} \label{friction_tensor:eq}
 \underline{\underline{\xi}} = \begin{pmatrix}
\xi_{\bot} & \xi_{\wedge} & 0 \\
-\xi_{\wedge} & \xi_{\bot} & 0 \\
0 & 0 & \xi_{\parallel}
\end{pmatrix}.
\end{equation}
The associated parallel, perpendicular, and transverse velocity correlation functions obtained from Eq. (\ref{magLangevin:eq}) are
\begin{subequations}
\label{Zpara:eq}
\begin{align} 
Z_{\parallel}(t) &= Z_{\parallel}(0) \exp{(-t \xi_{\parallel})} \label{Zperp:eq} \\
Z_{\bot}(t) &= Z_{\bot}(0) \exp{(-t \xi_{\bot})} \cos{\big[ (\omega_c - \xi_{\wedge})t \big]}
 \label{Zwedge:eq}\\
Z_{\wedge}(t) &= Z_{\bot}(0) \exp{(-t \xi_{\bot})} \sin{\big[ (\omega_c - \xi_{\wedge})t \big]}
\end{align}
\end{subequations}
and the resulting diffusion coefficients are
\begin{subequations}
\label{eq:diffusion_mag}
\begin{align}
\frac{D_{\parallel}}{\omega_p a^2} &= \frac{\omega_p}{3 \Gamma \xi_{\parallel}}
\label{Dpara_fit:eq} \\
\frac{D_{\bot}}{\omega_p a^2} &= \frac{\omega_p}{3 \Gamma} \bigg[ \frac{\xi_{\bot}}{\xi_{\bot}^2 + (\omega_c - \xi_{\wedge})^2} \bigg]
\label{Dperp_fit:eq} \\
\frac{D_{\wedge}}{\omega_p a^2} &= \frac{\omega_p}{3 \Gamma} \bigg[ \frac{\omega_c - \xi_{\wedge}}{\xi_{\bot}^2 + (\omega_c - \xi_{\wedge})^2} \bigg].
\label{Dwedge_fit:eq}
\end{align}
\end{subequations}
The friction coefficients were obtained by fitting the simulated velocity autocorrelation functions with Eq.~(\ref{Zpara:eq}) and the diffusion coefficients obtained from using the results in Eq.~(\ref{eq:diffusion_mag}).
To reduce the accumulation of errors, the fit was applied to early times ($\approx [0,25] \omega_p^{-1}$). 

The relaxation time approximation assumes correlations between particles are weak. 
Figure~\ref{fig:mag_g1_VACF} shows that although this is an excellent approximation at small $\Gamma$ when $\beta \ll 1$, it cannot capture the correlation caused by strong magnetization at the same $\Gamma$ value. 
This can be seen in both the parallel and perpendicular cumulative integrals in Fig.~\ref{fig:mag_g1_VACF}. 
It is particularly stark in the perpendicular direction, where the $Z_\perp$ predicted by the relaxation time approximation looks nearly indistinguishable from the MD simulations. 
However, there is a slight phase-shift which, when integrated, leads to very large differences in the cumulative integral, and therefore, the predicted diffusion coefficient. 
It is expected that the primary cause of the failure of the relaxation time model is the increased correlation scale introduced by strong magnetization. 
However, it should also be noted that it has recently been shown that the perpendicular friction force ($\xi_\perp$) in magnetized plasma depends on the angle between the particle's velocity and the magnetic field, and therefore cannot be characterized by a constant coefficient even in the low-speed limit.~\cite{Baalrud:Fx}  
This may couple $Z_\parallel$ and $Z_\perp$, and is beyond the limitations of the model presented here. 

\subsection{Correlation time\label{sec:time}}

\begin{figure}
\includegraphics[width=8.5cm]{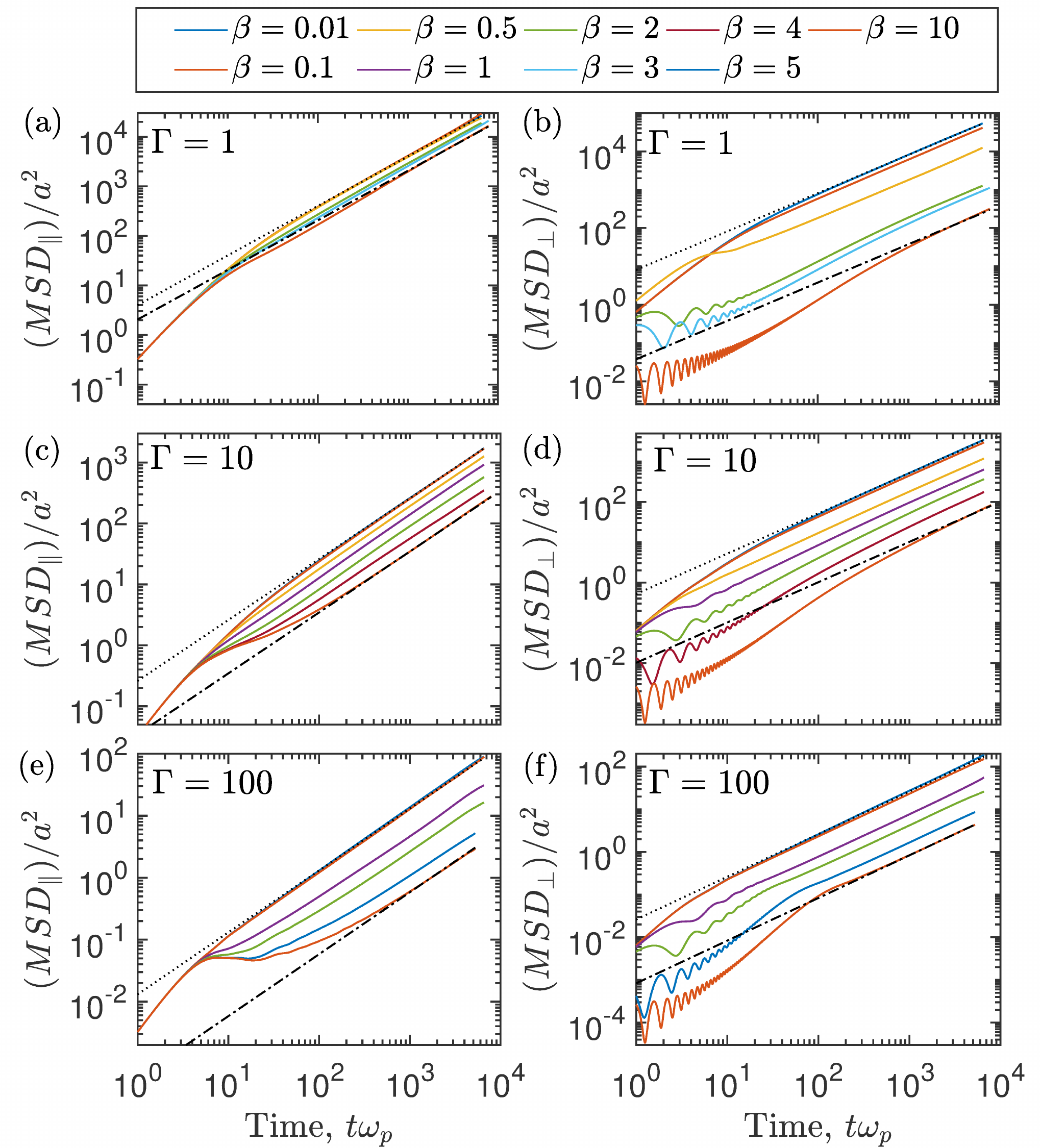}
\caption{Mean-squared-displacement (MSD) in the $\parallel$ and $\bot$ directions at several magnetic field strengths, $\beta$.  The dashed and dash-dotted lines are linear fits to the $\beta=0.01$ and $10$ data.}
\label{fig:MSD_beta}
\end{figure}  

The simultaneous influence of large momentum transfer collisions at the ends of the collision cylinder, and the continuous drifting aspect of the motion, leads to a decay of the velocity autocorrelation function that is characterized by two timescales. 
In a strongly magnetized plasma, the timescale associated with collisions from the ends of the collision cylinder can become very long; resulting in a longer time to reach the fluid (diffusive) regime.
This is illustrated in Fig.~\ref{fig:MSD_beta}, which shows the mean-square displacement in the parallel and perpendicular directions as a function of time over a wide range of magnetic field strengths ($\beta = 0.01-10$) for $\Gamma =1, 10$ and 100. 
In each case, there is an initial stage associated with the time for interactions to be established within the collision volume. 
In a weakly coupled plasma, this timescale is characterized by the plasma period because the collision diameter is the Debye length: $\tau \approx \lambda_{D}/v_T = \omega_p^{-1}$. 
In a strongly coupled plasma, collective interactions determine the interaction range, which also occur near the plasma period in the OCP. 
In each case shown in Fig.~\ref{fig:MSD_beta}, there is an initial slope that is steeper than linear in time associated with this regime that lasts for the first few plasma periods. 

Qualitative differences between the weakly and strongly magnetized regimes are observed at subsequent times, as the time it takes to reach a linear regime is much longer when $\beta > 1$, and the intermediate timescale is steeper than linear in the perpendicular direction and shallower than linear in the parallel direction.  
The extent of this intermediate timescale extends with increasing $\beta$. 
Considering panel (b) of Fig.~\ref{fig:MSD_beta}, the time to reach a linear regime when $\Gamma = 1$ and $\beta=10$ extends to more than 1000$\omega_p^{-1}$. 
Further evidence that this arrested decay is associated with the long-range parallel correlation is shown in panels (d) and (f), which illustrates that a linear regime is established sooner at higher $\Gamma$. 
As described in the previous section, and Fig.~\ref{fig:trajectories}, the parallel correlation scale shrinks at high $\Gamma$ due to the increased collisionality. 

\begin{figure*}
\includegraphics[width=16cm]{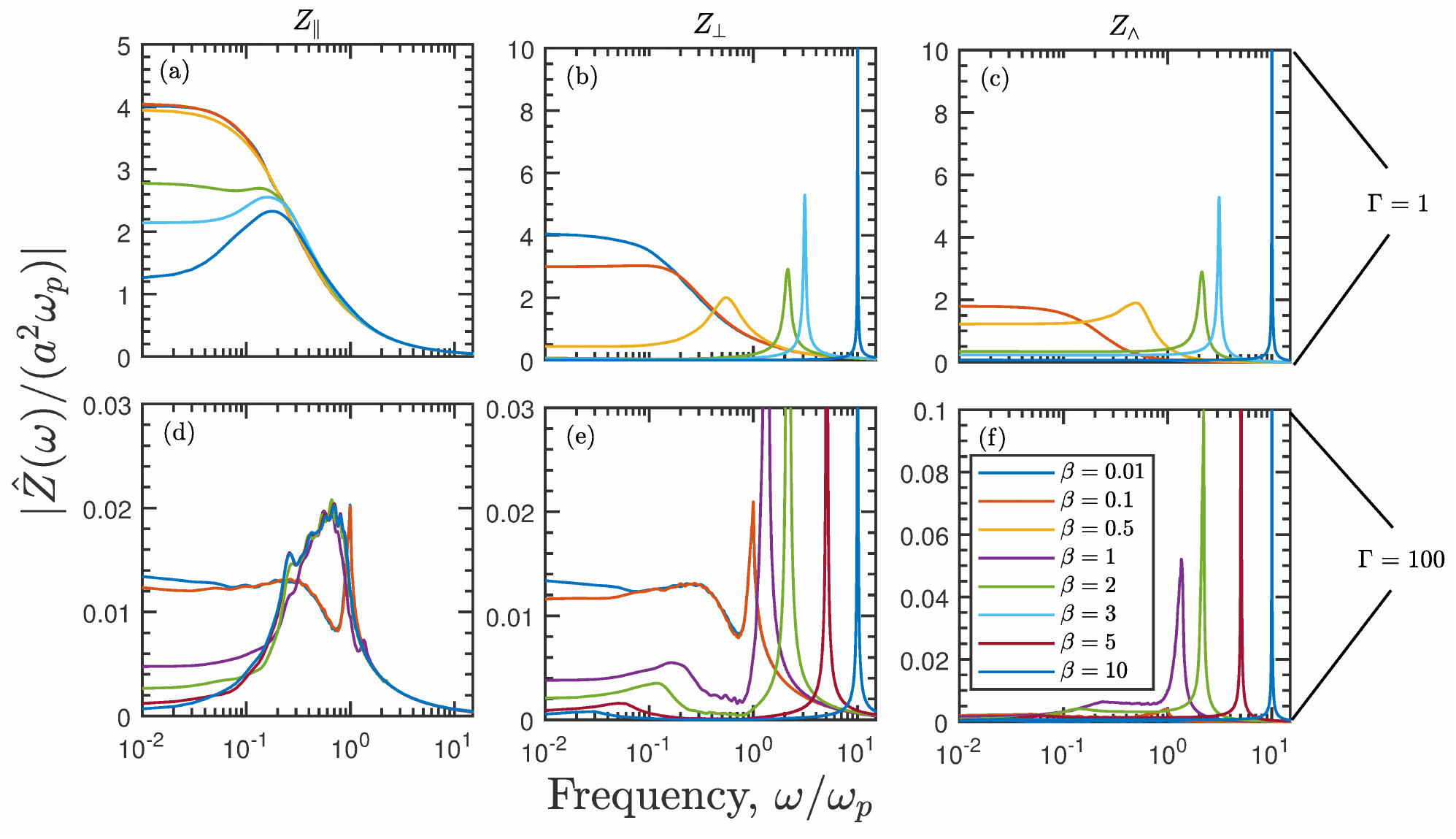}
\caption{The modulus of the Fourier transform, $| \hat{Z}(w) |$, of the parallel, perpendicular, and transverse velocity correlation function. In the legend, $\beta=0.5$ and $3$ are exclusive to $\Gamma=1$, and $\beta=1$ and $5$ are exclusive to $\Gamma=100$.}
\label{fig:FTs}
\end{figure*}

These two distinct timescales can be observed directly in the velocity autocorrelation functions and their cumulative integrals; see Fig.~\ref{fig:G10_b4}. 
In an unmagnetized plasma, the velocity autocorrelation function decays steadily over the collision timescale so that a single plateau is observed in the cummulative integral at a time directly following the time at which the velocity autocorrelation function has visibly decayed to near zero; see Fig.~\ref{fig:ZtandDt}. 
In contrast, two plateaus are observed at the strong magnetization conditions shown in Fig.~\ref{fig:G10_b4}. 
The first is associated with the time subsequent to the initial decay of the velocity autocorrelation function and may appear as though this value should correspond to the self-diffusion coefficient. 
However, if the simulation is run much longer in time, a second flatter plateau is observed much later, as shown in the right column of Fig.~\ref{fig:G10_b4}. 
Comparing the cumulative integral from Fig.~\ref{fig:G10_b4} to the mean square displacement from Fig.~\ref{fig:MSD_beta}(d) clearly shows that it is the late-time plateau that is associated with the diffusive regime.
The apparent plateau in the intermediate time interval is not quite flat, though it can appear so to within the simulation resolution. 

These time profiles indicate that strong magnetization arrests the approach to a fluid regime. 
If a simulation is not run long enough, or a measurement is taken too early in time, this can leave the impression that the perpendicular diffusion is superdiffusive. 
Superdiffusion refers to evolution in which the mean square displacement grows faster than linearly in time, such as in the interval from 100-1000$\omega_{p}^{-1}$ in Fig.~\ref{fig:MSD_beta}(b). 
However, we observe that after enough time has evolved to incorporate motion associated with the long-range parallel correlation, that a diffusive (linear in time) regime is reached.  

A very wide range of timescales must be resolved in the strongly magnetized regime, which presents a challenge for MD simulations. 
Figure~\ref{fig:FTs} shows the Fourier transform of the velocity correlation functions.
At zero frequency the value of $\hat{Z}(\omega)$ is equal to the diffusion coefficient.
At low magnetization ($\beta = 0.01$) and low coupling ($\Gamma = 1$) the spectrum in each direction is characterized by a single broad peak at low frequency, associated with the decay due to Coulomb collisions. 
The spectrum at low magnetization ($\beta = 0.01$) and strong coupling ($\Gamma = 100$) has a similar broad low frequency component, but also an additional sharp peak associated with the plasmon dispersion at $\omega = \omega_p$, and a broader peak at $\omega \approx 0.25 \omega_p$ that is associated with caging of particles by their nearest neighbors.~\cite{Donko:2002}

In contrast, multiple peaks are observed spanning a broad frequency range as $\beta$ increases; see Refs.~\onlinecite{OttPRL2012,OttPRE2013,KahlertPOP2013} for a thorough discussion of the oscillation spectrum of the magnetized OCP in 3D, and Refs.~\onlinecite{HartmannPRL2013,DzhumaulovaPRE2014} for 2D. 
The $\hat{Z}_{\bot}$ and $\hat{Z}_{\wedge}$ spectra show high-frequency peaks associated with the upper hybrid mode: a longitudinal wave that propagates perpendicular to the magnetic field at a frequency $\omega / \omega_p = \sqrt{1 + \beta^2}$. 
This high-frequency peak is near the gyrofrequency and sets the timestep requirement in the simulation.
In the parallel direction, a prominent peak develops at a frequency a few times lower than the plasma frequency (near $0.2\omega_p$ for $\Gamma=1$ and near $0.6\omega_p$ for $\Gamma=100$) as $\beta$ increases. 
This peak is likely due to the bouncing component of particle motion being extended by the magnetic field (i.e., the long-range parallel correlation), as shown in Fig.~\ref{fig:trajectories}, which occurs more slowly at $\Gamma = 1$ than at $\Gamma =100$. 
In both cases a broad low-frequency component also persists, which is likely associated with the continuous drifting aspect from long-range Coulomb collisions. 
In the perpendicular direction at large coupling ($\Gamma = 100$) the correlation peak associated with caging is observed to shift to lower frequency as $\beta$ increases, as expected from the parallel correlation associated with the formation of a collision cylinder (rather than sphere). 

\section{Conclusions\label{sec:conclusions}}

Molecular dynamics simulations demonstrated that strong magnetization ($\beta \gg 1$) gives rise to a long-range correlation associated with interparticle interactions. 
The physical mechanism is a channeling effect, as a magnetic field of sufficient strength confines particles to move largely in one dimension, scattering in nearly 180$^\circ$ interactions with particles on nearby magnetic field lines. 
The resulting interaction volume is characterized by an elongated collision cylinder, rather than the spherical geometry expected at weak magnetization. 
The elongation of the cylinder is reduced when the Coulomb coupling strength increases due to an increased collision rate. 
Influences of this correlation on the velocity autocorrelation function were observed to be reminiscent of what is observed at higher coupling strengths ($\Gamma \gg 1$) without strong magnetization, including negative correlations. 
Increased spatial correlations also translate to an increase in the time to reach a fluid plasma regime, in which the root mean square displacement reaches the expected linear in time scaling indicative of a diffusive process. 
 
Increases in space and time correlations have direct implications for experiments. 
Nonneutral plasmas routinely reach very strong magnetization conditions $(\beta \gg 1)$ and ultracold plasmas are beginning to reach this regime.~\cite{Beck:1992,Beck:1996,Dubin:1999,Roberts:2020}
One motivation for these experiments is to perform sensitive tests of plasma transport properties. 
Many of these experiments include on the order of $10^4-10^8$ charged particles in total. 
Our work indicates that the number of particles required for these experiments to represent a plasma transport regime when $\beta \gg 1$ is large enough to influence the outcome of previous and ongoing experiments.  
For example, our work indicates that a common condition of $\Gamma \approx 0.1$ and $\beta \approx 10$ may require a system size in excess of $10^8$ particles in order to represent a plasma. 



 




\section{Data Availability} 

The data that support the findings of this study are available from the corresponding author upon reasonable request.

\begin{acknowledgments}
The authors thank Dr.~Jerome Daligault for supplying the MD code used in this work, for training in use of the code, and for assistance in the data analysis and interpretation. 
This material is based upon work supported by the Air Force Office of Scientific Research under award number FA9550-16-1-0221 and by the National Science Foundation under award number PHY-1453736.  It use the Extreme Science and Engineering Discovery Environment (XSEDE) \cite{xsede}, which is supported by NSF  Grant No. ACI-1548562, under Project Award No. PHYS-150018. 
\end{acknowledgments}

\bibliography{references}

\begin{thebibliography}{54}%
\makeatletter
\providecommand \@ifxundefined [1]{%
 \@ifx{#1\undefined}
}%
\providecommand \@ifnum [1]{%
 \ifnum #1\expandafter \@firstoftwo
 \else \expandafter \@secondoftwo
 \fi
}%
\providecommand \@ifx [1]{%
 \ifx #1\expandafter \@firstoftwo
 \else \expandafter \@secondoftwo
 \fi
}%
\providecommand \natexlab [1]{#1}%
\providecommand \enquote  [1]{``#1''}%
\providecommand \bibnamefont  [1]{#1}%
\providecommand \bibfnamefont [1]{#1}%
\providecommand \citenamefont [1]{#1}%
\providecommand \href@noop [0]{\@secondoftwo}%
\providecommand \href [0]{\begingroup \@sanitize@url \@href}%
\providecommand \@href[1]{\@@startlink{#1}\@@href}%
\providecommand \@@href[1]{\endgroup#1\@@endlink}%
\providecommand \@sanitize@url [0]{\catcode `\\12\catcode `\$12\catcode
  `\&12\catcode `\#12\catcode `\^12\catcode `\_12\catcode `\%12\relax}%
\providecommand \@@startlink[1]{}%
\providecommand \@@endlink[0]{}%
\providecommand \url  [0]{\begingroup\@sanitize@url \@url }%
\providecommand \@url [1]{\endgroup\@href {#1}{\urlprefix }}%
\providecommand \urlprefix  [0]{URL }%
\providecommand \Eprint [0]{\href }%
\providecommand \doibase [0]{http://dx.doi.org/}%
\providecommand \selectlanguage [0]{\@gobble}%
\providecommand \bibinfo  [0]{\@secondoftwo}%
\providecommand \bibfield  [0]{\@secondoftwo}%
\providecommand \translation [1]{[#1]}%
\providecommand \BibitemOpen [0]{}%
\providecommand \bibitemStop [0]{}%
\providecommand \bibitemNoStop [0]{.\EOS\space}%
\providecommand \EOS [0]{\spacefactor3000\relax}%
\providecommand \BibitemShut  [1]{\csname bibitem#1\endcsname}%
\let\auto@bib@innerbib\@empty
\bibitem [{\citenamefont {Gurnett}\ and\ \citenamefont
  {Bhattacharjee}(2017)}]{Gurnett:Book}%
  \BibitemOpen
  \bibfield  {author} {\bibinfo {author} {\bibfnamefont {D.~A.}\ \bibnamefont
  {Gurnett}}\ and\ \bibinfo {author} {\bibfnamefont {A.}~\bibnamefont
  {Bhattacharjee}},\ }\href@noop {} {\emph {\bibinfo {title} {Introduction to
  Plasma Physics: With Space, Laboratory and Astrophysical Applications}}},\
  \bibinfo {edition} {2nd}\ ed.\ (\bibinfo  {publisher} {Cambridge University
  Press},\ \bibinfo {year} {2017})\BibitemShut {NoStop}%
\bibitem [{\citenamefont {Beck}, \citenamefont {Fajans},\ and\ \citenamefont
  {Malmberg}(1992)}]{Beck:1992}%
  \BibitemOpen
  \bibfield  {author} {\bibinfo {author} {\bibfnamefont {B.~R.}\ \bibnamefont
  {Beck}}, \bibinfo {author} {\bibfnamefont {J.}~\bibnamefont {Fajans}}, \ and\
  \bibinfo {author} {\bibfnamefont {J.~H.}\ \bibnamefont {Malmberg}},\
  }\href@noop {} {\bibfield  {journal} {\bibinfo  {journal} {Phys. Rev. Lett.}\
  }\textbf {\bibinfo {volume} {68}},\ \bibinfo {pages} {317} (\bibinfo {year}
  {1992})}\BibitemShut {NoStop}%
\bibitem [{\citenamefont {Beck}, \citenamefont {Fajans},\ and\ \citenamefont
  {Malmberg}(1996)}]{Beck:1996}%
  \BibitemOpen
  \bibfield  {author} {\bibinfo {author} {\bibfnamefont {B.~R.}\ \bibnamefont
  {Beck}}, \bibinfo {author} {\bibfnamefont {J.}~\bibnamefont {Fajans}}, \ and\
  \bibinfo {author} {\bibfnamefont {J.~H.}\ \bibnamefont {Malmberg}},\
  }\href@noop {} {\bibfield  {journal} {\bibinfo  {journal} {Phys. Plasmas}\
  }\textbf {\bibinfo {volume} {3}},\ \bibinfo {pages} {1250} (\bibinfo {year}
  {1996})}\BibitemShut {NoStop}%
\bibitem [{\citenamefont {Dubin}\ and\ \citenamefont
  {O'Neil}(1999)}]{Dubin:1999}%
  \BibitemOpen
  \bibfield  {author} {\bibinfo {author} {\bibfnamefont {D.~H.~E.}\
  \bibnamefont {Dubin}}\ and\ \bibinfo {author} {\bibfnamefont {T.~M.}\
  \bibnamefont {O'Neil}},\ }\href@noop {} {\bibfield  {journal} {\bibinfo
  {journal} {Rev. Mod. Phys.}\ }\textbf {\bibinfo {volume} {71}},\ \bibinfo
  {pages} {87} (\bibinfo {year} {1999})}\BibitemShut {NoStop}%
\bibitem [{\citenamefont {Hollmann}, \citenamefont {Anderegg},\ and\
  \citenamefont {Driscoll}(1999)}]{HollmannPRL1999}%
  \BibitemOpen
  \bibfield  {author} {\bibinfo {author} {\bibfnamefont {E.~M.}\ \bibnamefont
  {Hollmann}}, \bibinfo {author} {\bibfnamefont {F.}~\bibnamefont {Anderegg}},
  \ and\ \bibinfo {author} {\bibfnamefont {C.~F.}\ \bibnamefont {Driscoll}},\
  }\href {\doibase 10.1103/PhysRevLett.82.4839} {\bibfield  {journal} {\bibinfo
   {journal} {Phys. Rev. Lett.}\ }\textbf {\bibinfo {volume} {82}},\ \bibinfo
  {pages} {4839} (\bibinfo {year} {1999})}\BibitemShut {NoStop}%
\bibitem [{\citenamefont {Kriesel}\ and\ \citenamefont
  {Driscoll}(2001)}]{KrieselPRL2001}%
  \BibitemOpen
  \bibfield  {author} {\bibinfo {author} {\bibfnamefont {J.~M.}\ \bibnamefont
  {Kriesel}}\ and\ \bibinfo {author} {\bibfnamefont {C.~F.}\ \bibnamefont
  {Driscoll}},\ }\href {\doibase 10.1103/PhysRevLett.87.135003} {\bibfield
  {journal} {\bibinfo  {journal} {Phys. Rev. Lett.}\ }\textbf {\bibinfo
  {volume} {87}},\ \bibinfo {pages} {135003} (\bibinfo {year}
  {2001})}\BibitemShut {NoStop}%
\bibitem [{\citenamefont {Anderegg}\ \emph {et~al.}(2017)\citenamefont
  {Anderegg}, \citenamefont {Dubin}, \citenamefont {Affolter},\ and\
  \citenamefont {Driscoll}}]{AndereggPOP2017}%
  \BibitemOpen
  \bibfield  {author} {\bibinfo {author} {\bibfnamefont {F.}~\bibnamefont
  {Anderegg}}, \bibinfo {author} {\bibfnamefont {D.~H.~E.}\ \bibnamefont
  {Dubin}}, \bibinfo {author} {\bibfnamefont {M.}~\bibnamefont {Affolter}}, \
  and\ \bibinfo {author} {\bibfnamefont {C.~F.}\ \bibnamefont {Driscoll}},\
  }\href {\doibase 10.1063/1.4999350} {\bibfield  {journal} {\bibinfo
  {journal} {Physics of Plasmas}\ }\textbf {\bibinfo {volume} {24}},\ \bibinfo
  {pages} {092118} (\bibinfo {year} {2017})},\ \Eprint
  {http://arxiv.org/abs/https://doi.org/10.1063/1.4999350}
  {https://doi.org/10.1063/1.4999350} \BibitemShut {NoStop}%
\bibitem [{\citenamefont {Affolter}\ \emph {et~al.}(2016)\citenamefont
  {Affolter}, \citenamefont {Anderegg}, \citenamefont {Dubin},\ and\
  \citenamefont {Driscoll}}]{AffolterPRL2016}%
  \BibitemOpen
  \bibfield  {author} {\bibinfo {author} {\bibfnamefont {M.}~\bibnamefont
  {Affolter}}, \bibinfo {author} {\bibfnamefont {F.}~\bibnamefont {Anderegg}},
  \bibinfo {author} {\bibfnamefont {D.~H.~E.}\ \bibnamefont {Dubin}}, \ and\
  \bibinfo {author} {\bibfnamefont {C.~F.}\ \bibnamefont {Driscoll}},\ }\href
  {\doibase 10.1103/PhysRevLett.117.155001} {\bibfield  {journal} {\bibinfo
  {journal} {Phys. Rev. Lett.}\ }\textbf {\bibinfo {volume} {117}},\ \bibinfo
  {pages} {155001} (\bibinfo {year} {2016})}\BibitemShut {NoStop}%
\bibitem [{\citenamefont {Ott}\ and\ \citenamefont {Bonitz}(2011)}]{Ott:2011}%
  \BibitemOpen
  \bibfield  {author} {\bibinfo {author} {\bibfnamefont {T.}~\bibnamefont
  {Ott}}\ and\ \bibinfo {author} {\bibfnamefont {M.}~\bibnamefont {Bonitz}},\
  }\href@noop {} {\bibfield  {journal} {\bibinfo  {journal} {Phys. Rev. Lett.}\
  }\textbf {\bibinfo {volume} {107}} (\bibinfo {year} {2011})}\BibitemShut
  {NoStop}%
\bibitem [{\citenamefont {Ott}, \citenamefont {Bonitz},\ and\ \citenamefont
  {Donk\'o}(2015)}]{OttPRE2015}%
  \BibitemOpen
  \bibfield  {author} {\bibinfo {author} {\bibfnamefont {T.}~\bibnamefont
  {Ott}}, \bibinfo {author} {\bibfnamefont {M.}~\bibnamefont {Bonitz}}, \ and\
  \bibinfo {author} {\bibfnamefont {Z.}~\bibnamefont {Donk\'o}},\ }\href
  {\doibase 10.1103/PhysRevE.92.063105} {\bibfield  {journal} {\bibinfo
  {journal} {Phys. Rev. E}\ }\textbf {\bibinfo {volume} {92}},\ \bibinfo
  {pages} {063105} (\bibinfo {year} {2015})}\BibitemShut {NoStop}%
\bibitem [{\citenamefont {Ott}\ \emph {et~al.}(2017)\citenamefont {Ott},
  \citenamefont {Bonitz}, \citenamefont {Hartmann},\ and\ \citenamefont
  {Donk\'o}}]{OttPRE2017}%
  \BibitemOpen
  \bibfield  {author} {\bibinfo {author} {\bibfnamefont {T.}~\bibnamefont
  {Ott}}, \bibinfo {author} {\bibfnamefont {M.}~\bibnamefont {Bonitz}},
  \bibinfo {author} {\bibfnamefont {P.}~\bibnamefont {Hartmann}}, \ and\
  \bibinfo {author} {\bibfnamefont {Z.}~\bibnamefont {Donk\'o}},\ }\href
  {\doibase 10.1103/PhysRevE.95.013209} {\bibfield  {journal} {\bibinfo
  {journal} {Phys. Rev. E}\ }\textbf {\bibinfo {volume} {95}},\ \bibinfo
  {pages} {013209} (\bibinfo {year} {2017})}\BibitemShut {NoStop}%
\bibitem [{\citenamefont {Baalrud}\ and\ \citenamefont
  {Daligault}(2017)}]{Baalrud:regimes}%
  \BibitemOpen
  \bibfield  {author} {\bibinfo {author} {\bibfnamefont {S.~D.}\ \bibnamefont
  {Baalrud}}\ and\ \bibinfo {author} {\bibfnamefont {J.}~\bibnamefont
  {Daligault}},\ }\href@noop {} {\bibfield  {journal} {\bibinfo  {journal}
  {Phys. Rev. E}\ }\textbf {\bibinfo {volume} {96}} (\bibinfo {year}
  {2017})}\BibitemShut {NoStop}%
\bibitem [{\citenamefont {Scheiner}\ and\ \citenamefont
  {Baalrud}(2020)}]{ScheinerPRE2020}%
  \BibitemOpen
  \bibfield  {author} {\bibinfo {author} {\bibfnamefont {B.}~\bibnamefont
  {Scheiner}}\ and\ \bibinfo {author} {\bibfnamefont {S.~D.}\ \bibnamefont
  {Baalrud}},\ }\href {\doibase 10.1103/PhysRevE.102.063202} {\bibfield
  {journal} {\bibinfo  {journal} {Phys. Rev. E}\ }\textbf {\bibinfo {volume}
  {102}},\ \bibinfo {pages} {063202} (\bibinfo {year} {2020})}\BibitemShut
  {NoStop}%
\bibitem [{\citenamefont {Bernstein}\ \emph {et~al.}(2020)\citenamefont
  {Bernstein}, \citenamefont {Lafleur}, \citenamefont {Daligault},\ and\
  \citenamefont {Baalrud}}]{BernsteinPRE2020}%
  \BibitemOpen
  \bibfield  {author} {\bibinfo {author} {\bibfnamefont {D.~J.}\ \bibnamefont
  {Bernstein}}, \bibinfo {author} {\bibfnamefont {T.}~\bibnamefont {Lafleur}},
  \bibinfo {author} {\bibfnamefont {J.}~\bibnamefont {Daligault}}, \ and\
  \bibinfo {author} {\bibfnamefont {S.~D.}\ \bibnamefont {Baalrud}},\ }\href
  {\doibase 10.1103/PhysRevE.102.041201} {\bibfield  {journal} {\bibinfo
  {journal} {Phys. Rev. E}\ }\textbf {\bibinfo {volume} {102}},\ \bibinfo
  {pages} {041201} (\bibinfo {year} {2020})}\BibitemShut {NoStop}%
\bibitem [{\citenamefont {Danielson}\ \emph {et~al.}(2015)\citenamefont
  {Danielson}, \citenamefont {Dubin}, \citenamefont {Greaves},\ and\
  \citenamefont {Surko}}]{Danielson:2015}%
  \BibitemOpen
  \bibfield  {author} {\bibinfo {author} {\bibfnamefont {J.~R.}\ \bibnamefont
  {Danielson}}, \bibinfo {author} {\bibfnamefont {D.~H.~E.}\ \bibnamefont
  {Dubin}}, \bibinfo {author} {\bibfnamefont {R.~G.}\ \bibnamefont {Greaves}},
  \ and\ \bibinfo {author} {\bibfnamefont {C.~M.}\ \bibnamefont {Surko}},\
  }\href@noop {} {\bibfield  {journal} {\bibinfo  {journal} {Rev. Mod. Phys.}\
  }\textbf {\bibinfo {volume} {87}},\ \bibinfo {pages} {247} (\bibinfo {year}
  {2015})}\BibitemShut {NoStop}%
\bibitem [{\citenamefont {Roberts}, \citenamefont {Guthrie},\ and\
  \citenamefont {Jiang}(2020)}]{Roberts:2020}%
  \BibitemOpen
  \bibfield  {author} {\bibinfo {author} {\bibfnamefont {J.}~\bibnamefont
  {Roberts}}, \bibinfo {author} {\bibfnamefont {J.}~\bibnamefont {Guthrie}}, \
  and\ \bibinfo {author} {\bibfnamefont {P.}~\bibnamefont {Jiang}},\ }in\
  \href@noop {} {\emph {\bibinfo {booktitle} {62nd Annual Meeting of the APS
  Division of Plasma Physics}}}\ (\bibinfo  {publisher} {APS},\ \bibinfo
  {address} {Virtual Meeting},\ \bibinfo {year} {2020})\BibitemShut {NoStop}%
\bibitem [{\citenamefont {Thomas}, \citenamefont {Merlino},\ and\ \citenamefont
  {Rosenberg}(2012)}]{Thomas:2012}%
  \BibitemOpen
  \bibfield  {author} {\bibinfo {author} {\bibfnamefont {E.}~\bibnamefont
  {Thomas}}, \bibinfo {author} {\bibfnamefont {R.~L.}\ \bibnamefont {Merlino}},
  \ and\ \bibinfo {author} {\bibfnamefont {M.}~\bibnamefont {Rosenberg}},\
  }\href@noop {} {\bibfield  {journal} {\bibinfo  {journal} {Plasma Phys.
  Control. Fusion}\ }\textbf {\bibinfo {volume} {54}},\ \bibinfo {pages}
  {124034} (\bibinfo {year} {2012})}\BibitemShut {NoStop}%
\bibitem [{\citenamefont {Bonitz}\ \emph {et~al.}(2012)\citenamefont {Bonitz},
  \citenamefont {Kalhlert}, \citenamefont {Ott},\ and\ \citenamefont
  {Laowen}}]{Bonitz:2012}%
  \BibitemOpen
  \bibfield  {author} {\bibinfo {author} {\bibfnamefont {M.}~\bibnamefont
  {Bonitz}}, \bibinfo {author} {\bibfnamefont {H.}~\bibnamefont {Kalhlert}},
  \bibinfo {author} {\bibfnamefont {T.}~\bibnamefont {Ott}}, \ and\ \bibinfo
  {author} {\bibfnamefont {H.}~\bibnamefont {Laowen}},\ }\href@noop {}
  {\bibfield  {journal} {\bibinfo  {journal} {Plasma Sources Sci. Technol.}\
  }\textbf {\bibinfo {volume} {22}},\ \bibinfo {pages} {015007} (\bibinfo
  {year} {2012})}\BibitemShut {NoStop}%
\bibitem [{\citenamefont {Tadsen}, \citenamefont {Greiner},\ and\ \citenamefont
  {Piel}(2018)}]{TadsenPRE2018}%
  \BibitemOpen
  \bibfield  {author} {\bibinfo {author} {\bibfnamefont {B.}~\bibnamefont
  {Tadsen}}, \bibinfo {author} {\bibfnamefont {F.}~\bibnamefont {Greiner}}, \
  and\ \bibinfo {author} {\bibfnamefont {A.}~\bibnamefont {Piel}},\ }\href
  {\doibase 10.1103/PhysRevE.97.033203} {\bibfield  {journal} {\bibinfo
  {journal} {Phys. Rev. E}\ }\textbf {\bibinfo {volume} {97}},\ \bibinfo
  {pages} {033203} (\bibinfo {year} {2018})}\BibitemShut {NoStop}%
\bibitem [{\citenamefont {Hartmann}\ \emph {et~al.}(2019)\citenamefont
  {Hartmann}, \citenamefont {Reyes}, \citenamefont {Kostadinova}, \citenamefont
  {Matthews}, \citenamefont {Hyde}, \citenamefont {Masheyeva}, \citenamefont
  {Dzhumagulova}, \citenamefont {Ramazanov}, \citenamefont {Ott}, \citenamefont
  {K\"ahlert}, \citenamefont {Bonitz}, \citenamefont {Korolov},\ and\
  \citenamefont {Donk\'o}}]{HartmannPRE2019}%
  \BibitemOpen
  \bibfield  {author} {\bibinfo {author} {\bibfnamefont {P.}~\bibnamefont
  {Hartmann}}, \bibinfo {author} {\bibfnamefont {J.~C.}\ \bibnamefont {Reyes}},
  \bibinfo {author} {\bibfnamefont {E.~G.}\ \bibnamefont {Kostadinova}},
  \bibinfo {author} {\bibfnamefont {L.~S.}\ \bibnamefont {Matthews}}, \bibinfo
  {author} {\bibfnamefont {T.~W.}\ \bibnamefont {Hyde}}, \bibinfo {author}
  {\bibfnamefont {R.~U.}\ \bibnamefont {Masheyeva}}, \bibinfo {author}
  {\bibfnamefont {K.~N.}\ \bibnamefont {Dzhumagulova}}, \bibinfo {author}
  {\bibfnamefont {T.~S.}\ \bibnamefont {Ramazanov}}, \bibinfo {author}
  {\bibfnamefont {T.}~\bibnamefont {Ott}}, \bibinfo {author} {\bibfnamefont
  {H.}~\bibnamefont {K\"ahlert}}, \bibinfo {author} {\bibfnamefont
  {M.}~\bibnamefont {Bonitz}}, \bibinfo {author} {\bibfnamefont
  {I.}~\bibnamefont {Korolov}}, \ and\ \bibinfo {author} {\bibfnamefont
  {Z.}~\bibnamefont {Donk\'o}},\ }\href {\doibase 10.1103/PhysRevE.99.013203}
  {\bibfield  {journal} {\bibinfo  {journal} {Phys. Rev. E}\ }\textbf {\bibinfo
  {volume} {99}},\ \bibinfo {pages} {013203} (\bibinfo {year}
  {2019})}\BibitemShut {NoStop}%
\bibitem [{\citenamefont {Gotchev}\ \emph {et~al.}(2009)\citenamefont
  {Gotchev}, \citenamefont {Chang}, \citenamefont {Knauer}, \citenamefont
  {Meyerhofer}, \citenamefont {Polomarov}, \citenamefont {Frenje},
  \citenamefont {Li}, \citenamefont {Manuel}, \citenamefont {Petrasso},
  \citenamefont {Rygg}, \citenamefont {S\'eguin},\ and\ \citenamefont
  {Betti}}]{Gotchev:2009}%
  \BibitemOpen
  \bibfield  {author} {\bibinfo {author} {\bibfnamefont {O.~V.}\ \bibnamefont
  {Gotchev}}, \bibinfo {author} {\bibfnamefont {P.~Y.}\ \bibnamefont {Chang}},
  \bibinfo {author} {\bibfnamefont {J.~P.}\ \bibnamefont {Knauer}}, \bibinfo
  {author} {\bibfnamefont {D.~D.}\ \bibnamefont {Meyerhofer}}, \bibinfo
  {author} {\bibfnamefont {O.}~\bibnamefont {Polomarov}}, \bibinfo {author}
  {\bibfnamefont {J.}~\bibnamefont {Frenje}}, \bibinfo {author} {\bibfnamefont
  {C.~K.}\ \bibnamefont {Li}}, \bibinfo {author} {\bibfnamefont {M.~J.-E.}\
  \bibnamefont {Manuel}}, \bibinfo {author} {\bibfnamefont {R.~D.}\
  \bibnamefont {Petrasso}}, \bibinfo {author} {\bibfnamefont {J.~R.}\
  \bibnamefont {Rygg}}, \bibinfo {author} {\bibfnamefont {F.~H.}\ \bibnamefont
  {S\'eguin}}, \ and\ \bibinfo {author} {\bibfnamefont {R.}~\bibnamefont
  {Betti}},\ }\href@noop {} {\bibfield  {journal} {\bibinfo  {journal} {Phys.
  Rev. Lett.}\ }\textbf {\bibinfo {volume} {103}},\ \bibinfo {pages} {215004}
  (\bibinfo {year} {2009})}\BibitemShut {NoStop}%
\bibitem [{\citenamefont {Hohenberger}\ \emph {et~al.}(2012)\citenamefont
  {Hohenberger}, \citenamefont {Chang}, \citenamefont {Fiksel}, \citenamefont
  {Knauer}, \citenamefont {Betti}, \citenamefont {Marshall}, \citenamefont
  {Meyerhofer}, \citenamefont {Séguin},\ and\ \citenamefont
  {Petrasso}}]{Hohenberger:2012}%
  \BibitemOpen
  \bibfield  {author} {\bibinfo {author} {\bibfnamefont {M.}~\bibnamefont
  {Hohenberger}}, \bibinfo {author} {\bibfnamefont {P.-Y.}\ \bibnamefont
  {Chang}}, \bibinfo {author} {\bibfnamefont {G.}~\bibnamefont {Fiksel}},
  \bibinfo {author} {\bibfnamefont {J.~P.}\ \bibnamefont {Knauer}}, \bibinfo
  {author} {\bibfnamefont {R.}~\bibnamefont {Betti}}, \bibinfo {author}
  {\bibfnamefont {F.~J.}\ \bibnamefont {Marshall}}, \bibinfo {author}
  {\bibfnamefont {D.~D.}\ \bibnamefont {Meyerhofer}}, \bibinfo {author}
  {\bibfnamefont {F.~H.}\ \bibnamefont {Séguin}}, \ and\ \bibinfo {author}
  {\bibfnamefont {R.~D.}\ \bibnamefont {Petrasso}},\ }\href@noop {} {\bibfield
  {journal} {\bibinfo  {journal} {Phys. Plasmas}\ }\textbf {\bibinfo {volume}
  {19}},\ \bibinfo {pages} {056306} (\bibinfo {year} {2012})}\BibitemShut
  {NoStop}%
\bibitem [{\citenamefont {Gomez}\ \emph {et~al.}(2014)\citenamefont {Gomez},
  \citenamefont {Slutz}, \citenamefont {Sefkow}, \citenamefont {Sinars},
  \citenamefont {Hahn}, \citenamefont {Hansen}, \citenamefont {Harding},
  \citenamefont {Knapp}, \citenamefont {Schmit}, \citenamefont {Jennings},
  \citenamefont {Awe}, \citenamefont {Geissel}, \citenamefont {Rovang},
  \citenamefont {Chandler}, \citenamefont {Cooper}, \citenamefont {Cuneo},
  \citenamefont {Harvey-Thompson}, \citenamefont {Herrmann}, \citenamefont
  {Hess}, \citenamefont {Johns}, \citenamefont {Lamppa}, \citenamefont
  {Martin}, \citenamefont {McBride}, \citenamefont {Peterson}, \citenamefont
  {Porter}, \citenamefont {Robertson}, \citenamefont {Rochau}, \citenamefont
  {Ruiz}, \citenamefont {Savage}, \citenamefont {Smith}, \citenamefont
  {Stygar},\ and\ \citenamefont {Vesey}}]{Gomez:2014}%
  \BibitemOpen
  \bibfield  {author} {\bibinfo {author} {\bibfnamefont {M.~R.}\ \bibnamefont
  {Gomez}}, \bibinfo {author} {\bibfnamefont {S.~A.}\ \bibnamefont {Slutz}},
  \bibinfo {author} {\bibfnamefont {A.~B.}\ \bibnamefont {Sefkow}}, \bibinfo
  {author} {\bibfnamefont {D.~B.}\ \bibnamefont {Sinars}}, \bibinfo {author}
  {\bibfnamefont {K.~D.}\ \bibnamefont {Hahn}}, \bibinfo {author}
  {\bibfnamefont {S.~B.}\ \bibnamefont {Hansen}}, \bibinfo {author}
  {\bibfnamefont {E.~C.}\ \bibnamefont {Harding}}, \bibinfo {author}
  {\bibfnamefont {P.~F.}\ \bibnamefont {Knapp}}, \bibinfo {author}
  {\bibfnamefont {P.~F.}\ \bibnamefont {Schmit}}, \bibinfo {author}
  {\bibfnamefont {C.~A.}\ \bibnamefont {Jennings}}, \bibinfo {author}
  {\bibfnamefont {T.~J.}\ \bibnamefont {Awe}}, \bibinfo {author} {\bibfnamefont
  {M.}~\bibnamefont {Geissel}}, \bibinfo {author} {\bibfnamefont {D.~C.}\
  \bibnamefont {Rovang}}, \bibinfo {author} {\bibfnamefont {G.~A.}\
  \bibnamefont {Chandler}}, \bibinfo {author} {\bibfnamefont {G.~W.}\
  \bibnamefont {Cooper}}, \bibinfo {author} {\bibfnamefont {M.~E.}\
  \bibnamefont {Cuneo}}, \bibinfo {author} {\bibfnamefont {A.~J.}\ \bibnamefont
  {Harvey-Thompson}}, \bibinfo {author} {\bibfnamefont {M.~C.}\ \bibnamefont
  {Herrmann}}, \bibinfo {author} {\bibfnamefont {M.~H.}\ \bibnamefont {Hess}},
  \bibinfo {author} {\bibfnamefont {O.}~\bibnamefont {Johns}}, \bibinfo
  {author} {\bibfnamefont {D.~C.}\ \bibnamefont {Lamppa}}, \bibinfo {author}
  {\bibfnamefont {M.~R.}\ \bibnamefont {Martin}}, \bibinfo {author}
  {\bibfnamefont {R.~D.}\ \bibnamefont {McBride}}, \bibinfo {author}
  {\bibfnamefont {K.~J.}\ \bibnamefont {Peterson}}, \bibinfo {author}
  {\bibfnamefont {J.~L.}\ \bibnamefont {Porter}}, \bibinfo {author}
  {\bibfnamefont {G.~K.}\ \bibnamefont {Robertson}}, \bibinfo {author}
  {\bibfnamefont {G.~A.}\ \bibnamefont {Rochau}}, \bibinfo {author}
  {\bibfnamefont {C.~L.}\ \bibnamefont {Ruiz}}, \bibinfo {author}
  {\bibfnamefont {M.~E.}\ \bibnamefont {Savage}}, \bibinfo {author}
  {\bibfnamefont {I.~C.}\ \bibnamefont {Smith}}, \bibinfo {author}
  {\bibfnamefont {W.~A.}\ \bibnamefont {Stygar}}, \ and\ \bibinfo {author}
  {\bibfnamefont {R.~A.}\ \bibnamefont {Vesey}},\ }\href@noop {} {\bibfield
  {journal} {\bibinfo  {journal} {Phys. Rev. Lett.}\ }\textbf {\bibinfo
  {volume} {113}},\ \bibinfo {pages} {155003} (\bibinfo {year}
  {2014})}\BibitemShut {NoStop}%
\bibitem [{\citenamefont {Shimada}\ \emph {et~al.}(2007)\citenamefont
  {Shimada}, \citenamefont {Campbell}, \citenamefont {Mukhovatov},
  \citenamefont {Fujiwara}, \citenamefont {Kirneva}, \citenamefont {Lackner},
  \citenamefont {Nagami}, \citenamefont {Pustovitov}, \citenamefont {Uckan},
  \citenamefont {Wesley}, \citenamefont {Asakura}, \citenamefont {Costley},
  \citenamefont {Donn{\'{e}}}, \citenamefont {Doyle}, \citenamefont {Fasoli},
  \citenamefont {Gormezano}, \citenamefont {Gribov}, \citenamefont {Gruber},
  \citenamefont {Hender}, \citenamefont {Houlberg}, \citenamefont {Ide},
  \citenamefont {Kamada}, \citenamefont {Leonard}, \citenamefont {Lipschultz},
  \citenamefont {Loarte}, \citenamefont {Miyamoto}, \citenamefont {Mukhovatov},
  \citenamefont {Osborne}, \citenamefont {Polevoi},\ and\ \citenamefont
  {Sips}}]{Shimada:2007}%
  \BibitemOpen
  \bibfield  {author} {\bibinfo {author} {\bibfnamefont {M.}~\bibnamefont
  {Shimada}}, \bibinfo {author} {\bibfnamefont {D.}~\bibnamefont {Campbell}},
  \bibinfo {author} {\bibfnamefont {V.}~\bibnamefont {Mukhovatov}}, \bibinfo
  {author} {\bibfnamefont {M.}~\bibnamefont {Fujiwara}}, \bibinfo {author}
  {\bibfnamefont {N.}~\bibnamefont {Kirneva}}, \bibinfo {author} {\bibfnamefont
  {K.}~\bibnamefont {Lackner}}, \bibinfo {author} {\bibfnamefont
  {M.}~\bibnamefont {Nagami}}, \bibinfo {author} {\bibfnamefont
  {V.}~\bibnamefont {Pustovitov}}, \bibinfo {author} {\bibfnamefont
  {N.}~\bibnamefont {Uckan}}, \bibinfo {author} {\bibfnamefont
  {J.}~\bibnamefont {Wesley}}, \bibinfo {author} {\bibfnamefont
  {N.}~\bibnamefont {Asakura}}, \bibinfo {author} {\bibfnamefont
  {A.}~\bibnamefont {Costley}}, \bibinfo {author} {\bibfnamefont
  {A.}~\bibnamefont {Donn{\'{e}}}}, \bibinfo {author} {\bibfnamefont
  {E.}~\bibnamefont {Doyle}}, \bibinfo {author} {\bibfnamefont
  {A.}~\bibnamefont {Fasoli}}, \bibinfo {author} {\bibfnamefont
  {C.}~\bibnamefont {Gormezano}}, \bibinfo {author} {\bibfnamefont
  {Y.}~\bibnamefont {Gribov}}, \bibinfo {author} {\bibfnamefont
  {O.}~\bibnamefont {Gruber}}, \bibinfo {author} {\bibfnamefont
  {T.}~\bibnamefont {Hender}}, \bibinfo {author} {\bibfnamefont
  {W.}~\bibnamefont {Houlberg}}, \bibinfo {author} {\bibfnamefont
  {S.}~\bibnamefont {Ide}}, \bibinfo {author} {\bibfnamefont {Y.}~\bibnamefont
  {Kamada}}, \bibinfo {author} {\bibfnamefont {A.}~\bibnamefont {Leonard}},
  \bibinfo {author} {\bibfnamefont {B.}~\bibnamefont {Lipschultz}}, \bibinfo
  {author} {\bibfnamefont {A.}~\bibnamefont {Loarte}}, \bibinfo {author}
  {\bibfnamefont {K.}~\bibnamefont {Miyamoto}}, \bibinfo {author}
  {\bibfnamefont {V.}~\bibnamefont {Mukhovatov}}, \bibinfo {author}
  {\bibfnamefont {T.}~\bibnamefont {Osborne}}, \bibinfo {author} {\bibfnamefont
  {A.}~\bibnamefont {Polevoi}}, \ and\ \bibinfo {author} {\bibfnamefont
  {A.}~\bibnamefont {Sips}},\ }\href@noop {} {\bibfield  {journal} {\bibinfo
  {journal} {Nucl. Fusion}\ }\textbf {\bibinfo {volume} {47}},\ \bibinfo
  {pages} {S1} (\bibinfo {year} {2007})}\BibitemShut {NoStop}%
\bibitem [{\citenamefont {Paz-Soldan}\ \emph {et~al.}(2014)\citenamefont
  {Paz-Soldan}, \citenamefont {Eidietis}, \citenamefont {Granetz},
  \citenamefont {Hollmann}, \citenamefont {Moyer}, \citenamefont {Wesley},
  \citenamefont {Zhang}, \citenamefont {Austin}, \citenamefont {Crocker},
  \citenamefont {Wingen},\ and\ \citenamefont {Zhu}}]{Paz-Soldan:2014}%
  \BibitemOpen
  \bibfield  {author} {\bibinfo {author} {\bibfnamefont {C.}~\bibnamefont
  {Paz-Soldan}}, \bibinfo {author} {\bibfnamefont {N.~W.}\ \bibnamefont
  {Eidietis}}, \bibinfo {author} {\bibfnamefont {R.}~\bibnamefont {Granetz}},
  \bibinfo {author} {\bibfnamefont {E.~M.}\ \bibnamefont {Hollmann}}, \bibinfo
  {author} {\bibfnamefont {R.~A.}\ \bibnamefont {Moyer}}, \bibinfo {author}
  {\bibfnamefont {J.~C.}\ \bibnamefont {Wesley}}, \bibinfo {author}
  {\bibfnamefont {J.}~\bibnamefont {Zhang}}, \bibinfo {author} {\bibfnamefont
  {M.~E.}\ \bibnamefont {Austin}}, \bibinfo {author} {\bibfnamefont {N.~A.}\
  \bibnamefont {Crocker}}, \bibinfo {author} {\bibfnamefont {A.}~\bibnamefont
  {Wingen}}, \ and\ \bibinfo {author} {\bibfnamefont {Y.}~\bibnamefont {Zhu}},\
  }\href@noop {} {\bibfield  {journal} {\bibinfo  {journal} {Phys. Plasmas}\
  }\textbf {\bibinfo {volume} {21}},\ \bibinfo {pages} {022514} (\bibinfo
  {year} {2014})}\BibitemShut {NoStop}%
\bibitem [{\citenamefont {Creely}\ \emph {et~al.}(2020)\citenamefont {Creely},
  \citenamefont {Greenwald}, \citenamefont {Ballinger}, \citenamefont
  {Brunner}, \citenamefont {Canik}, \citenamefont {Doody}, \citenamefont
  {Fülöp}, \citenamefont {Garnier}, \citenamefont {Granetz}, \citenamefont
  {Gray},\ and\ \citenamefont {et~al.}}]{Creely:2020}%
  \BibitemOpen
  \bibfield  {author} {\bibinfo {author} {\bibfnamefont {A.~J.}\ \bibnamefont
  {Creely}}, \bibinfo {author} {\bibfnamefont {M.~J.}\ \bibnamefont
  {Greenwald}}, \bibinfo {author} {\bibfnamefont {S.~B.}\ \bibnamefont
  {Ballinger}}, \bibinfo {author} {\bibfnamefont {D.}~\bibnamefont {Brunner}},
  \bibinfo {author} {\bibfnamefont {J.}~\bibnamefont {Canik}}, \bibinfo
  {author} {\bibfnamefont {J.}~\bibnamefont {Doody}}, \bibinfo {author}
  {\bibfnamefont {T.}~\bibnamefont {Fülöp}}, \bibinfo {author} {\bibfnamefont
  {D.~T.}\ \bibnamefont {Garnier}}, \bibinfo {author} {\bibfnamefont
  {R.}~\bibnamefont {Granetz}}, \bibinfo {author} {\bibfnamefont {T.~K.}\
  \bibnamefont {Gray}}, \ and\ \bibinfo {author} {\bibnamefont {et~al.}},\
  }\href@noop {} {\bibfield  {journal} {\bibinfo  {journal} {J. Plasma Phys.}\
  }\textbf {\bibinfo {volume} {86}},\ \bibinfo {pages} {865860502} (\bibinfo
  {year} {2020})}\BibitemShut {NoStop}%
\bibitem [{\citenamefont {{Khurana}}\ \emph {et~al.}(2004)\citenamefont
  {{Khurana}}, \citenamefont {{Kivelson}}, \citenamefont {{Vasyliunas}},
  \citenamefont {{Krupp}}, \citenamefont {{Woch}}, \citenamefont {{Lagg}},
  \citenamefont {{Mauk}},\ and\ \citenamefont {{Kurth}}}]{Khurana:2004}%
  \BibitemOpen
  \bibfield  {author} {\bibinfo {author} {\bibfnamefont {K.~K.}\ \bibnamefont
  {{Khurana}}}, \bibinfo {author} {\bibfnamefont {M.~G.}\ \bibnamefont
  {{Kivelson}}}, \bibinfo {author} {\bibfnamefont {V.~M.}\ \bibnamefont
  {{Vasyliunas}}}, \bibinfo {author} {\bibfnamefont {N.}~\bibnamefont
  {{Krupp}}}, \bibinfo {author} {\bibfnamefont {J.}~\bibnamefont {{Woch}}},
  \bibinfo {author} {\bibfnamefont {A.}~\bibnamefont {{Lagg}}}, \bibinfo
  {author} {\bibfnamefont {B.~H.}\ \bibnamefont {{Mauk}}}, \ and\ \bibinfo
  {author} {\bibfnamefont {W.~S.}\ \bibnamefont {{Kurth}}},\ }in\ \href@noop {}
  {\emph {\bibinfo {booktitle} {Jupiter: The Planet, Satellites and
  Magnetosphere}}},\ Vol.~\bibinfo {volume} {1},\ \bibinfo {editor} {edited by\
  \bibinfo {editor} {\bibfnamefont {F.}~\bibnamefont {{Bagenal}}}, \bibinfo
  {editor} {\bibfnamefont {T.~E.}\ \bibnamefont {{Dowling}}}, \ and\ \bibinfo
  {editor} {\bibfnamefont {W.~B.}\ \bibnamefont {{McKinnon}}}}\ (\bibinfo
  {publisher} {Cambridge University Press},\ \bibinfo {address} {New York},\
  \bibinfo {year} {2004})\ Chap.~\bibinfo {chapter} {24}, pp.\ \bibinfo {pages}
  {593--616}\BibitemShut {NoStop}%
\bibitem [{\citenamefont {Harding}\ and\ \citenamefont
  {Lai}(2006)}]{Harding:2006}%
  \BibitemOpen
  \bibfield  {author} {\bibinfo {author} {\bibfnamefont {A.~K.}\ \bibnamefont
  {Harding}}\ and\ \bibinfo {author} {\bibfnamefont {D.}~\bibnamefont {Lai}},\
  }\href@noop {} {\bibfield  {journal} {\bibinfo  {journal} {Rep. Prog. Phys.}\
  }\textbf {\bibinfo {volume} {69}},\ \bibinfo {pages} {2631} (\bibinfo {year}
  {2006})}\BibitemShut {NoStop}%
\bibitem [{\citenamefont {Hockney}\ and\ \citenamefont
  {Eastwood}(1981)}]{Hockney:MD}%
  \BibitemOpen
  \bibfield  {author} {\bibinfo {author} {\bibfnamefont {R.~W.}\ \bibnamefont
  {Hockney}}\ and\ \bibinfo {author} {\bibfnamefont {J.~W.}\ \bibnamefont
  {Eastwood}},\ }\href@noop {} {\emph {\bibinfo {title} {Computer Simulation
  Using Particles}}}\ (\bibinfo  {publisher} {McGraw-Hill Inc.},\ \bibinfo
  {year} {1981})\BibitemShut {NoStop}%
\bibitem [{\citenamefont {Tiwari}\ and\ \citenamefont
  {Baalrud}(2018)}]{TiwariPOP2018}%
  \BibitemOpen
  \bibfield  {author} {\bibinfo {author} {\bibfnamefont {S.~K.}\ \bibnamefont
  {Tiwari}}\ and\ \bibinfo {author} {\bibfnamefont {S.~D.}\ \bibnamefont
  {Baalrud}},\ }\href {\doibase 10.1063/1.5013320} {\bibfield  {journal}
  {\bibinfo  {journal} {Physics of Plasmas}\ }\textbf {\bibinfo {volume}
  {25}},\ \bibinfo {pages} {013511} (\bibinfo {year} {2018})},\ \Eprint
  {http://arxiv.org/abs/https://doi.org/10.1063/1.5013320}
  {https://doi.org/10.1063/1.5013320} \BibitemShut {NoStop}%
\bibitem [{\citenamefont {Gorman}\ \emph {et~al.}(2020)\citenamefont {Gorman},
  \citenamefont {Warrens}, \citenamefont {Bradshaw},\ and\ \citenamefont
  {Killian}}]{Gorman:2020}%
  \BibitemOpen
  \bibfield  {author} {\bibinfo {author} {\bibfnamefont {G.~M.}\ \bibnamefont
  {Gorman}}, \bibinfo {author} {\bibfnamefont {M.~K.}\ \bibnamefont {Warrens}},
  \bibinfo {author} {\bibfnamefont {S.~J.}\ \bibnamefont {Bradshaw}}, \ and\
  \bibinfo {author} {\bibfnamefont {T.~C.}\ \bibnamefont {Killian}},\
  }\href@noop {} {\bibfield  {journal} {\bibinfo  {journal} {arXiv}\ }
  (\bibinfo {year} {2020})}\BibitemShut {NoStop}%
\bibitem [{\citenamefont {Isaev}\ and\ \citenamefont
  {Gavriliuk}(2017)}]{IsaevJPB2018}%
  \BibitemOpen
  \bibfield  {author} {\bibinfo {author} {\bibfnamefont {I.~L.}\ \bibnamefont
  {Isaev}}\ and\ \bibinfo {author} {\bibfnamefont {A.~P.}\ \bibnamefont
  {Gavriliuk}},\ }\href {\doibase 10.1088/1361-6455/aa9b98} {\bibfield
  {journal} {\bibinfo  {journal} {Journal of Physics B: Atomic, Molecular and
  Optical Physics}\ }\textbf {\bibinfo {volume} {51}},\ \bibinfo {pages}
  {025701} (\bibinfo {year} {2017})}\BibitemShut {NoStop}%
\bibitem [{\citenamefont {Hansen}\ and\ \citenamefont
  {McDonald}(2013)}]{Hansen:2013}%
  \BibitemOpen
  \bibfield  {author} {\bibinfo {author} {\bibfnamefont {J.~P.}\ \bibnamefont
  {Hansen}}\ and\ \bibinfo {author} {\bibfnamefont {I.~R.}\ \bibnamefont
  {McDonald}},\ }\href@noop {} {\emph {\bibinfo {title} {Theory of Simple
  Liquids}}},\ \bibinfo {edition} {4th}\ ed.\ (\bibinfo  {publisher} {Academic
  Press},\ \bibinfo {address} {Oxford},\ \bibinfo {year} {2013})\BibitemShut
  {NoStop}%
\bibitem [{\citenamefont {Feng}\ \emph {et~al.}(2014)\citenamefont {Feng},
  \citenamefont {Goree}, \citenamefont {Liu}, \citenamefont {Intrator},\ and\
  \citenamefont {Murillo}}]{Feng:2014}%
  \BibitemOpen
  \bibfield  {author} {\bibinfo {author} {\bibfnamefont {Y.}~\bibnamefont
  {Feng}}, \bibinfo {author} {\bibfnamefont {J.}~\bibnamefont {Goree}},
  \bibinfo {author} {\bibfnamefont {B.}~\bibnamefont {Liu}}, \bibinfo {author}
  {\bibfnamefont {T.~P.}\ \bibnamefont {Intrator}}, \ and\ \bibinfo {author}
  {\bibfnamefont {M.~S.}\ \bibnamefont {Murillo}},\ }\href@noop {} {\bibfield
  {journal} {\bibinfo  {journal} {Phys. Rev. E}\ }\textbf {\bibinfo {volume}
  {90}} (\bibinfo {year} {2014})}\BibitemShut {NoStop}%
\bibitem [{\citenamefont {Frenkel}\ and\ \citenamefont
  {Smit}(2002)}]{Frenkel:MD}%
  \BibitemOpen
  \bibfield  {author} {\bibinfo {author} {\bibfnamefont {D.}~\bibnamefont
  {Frenkel}}\ and\ \bibinfo {author} {\bibfnamefont {B.}~\bibnamefont {Smit}},\
  }\href@noop {} {\emph {\bibinfo {title} {Understanding Molecular Simulation:
  From Algorithms to Applications}}}\ (\bibinfo  {publisher} {Academic Press},\
  \bibinfo {address} {San Diego, CA},\ \bibinfo {year} {2002})\BibitemShut
  {NoStop}%
\bibitem [{\citenamefont {Spreiter}\ and\ \citenamefont
  {Walter}(1999)}]{Spreiter:1999}%
  \BibitemOpen
  \bibfield  {author} {\bibinfo {author} {\bibfnamefont {Q.}~\bibnamefont
  {Spreiter}}\ and\ \bibinfo {author} {\bibfnamefont {M.}~\bibnamefont
  {Walter}},\ }\href@noop {} {\bibfield  {journal} {\bibinfo  {journal} {J.
  Comput. Phys.}\ }\textbf {\bibinfo {volume} {152}},\ \bibinfo {pages} {102 }
  (\bibinfo {year} {1999})}\BibitemShut {NoStop}%
\bibitem [{\citenamefont {Pathria}\ and\ \citenamefont
  {Beale}(2011)}]{Pathria:2011}%
  \BibitemOpen
  \bibfield  {author} {\bibinfo {author} {\bibfnamefont {R.}~\bibnamefont
  {Pathria}}\ and\ \bibinfo {author} {\bibfnamefont {P.~D.}\ \bibnamefont
  {Beale}},\ }\href@noop {} {\emph {\bibinfo {title} {Statistical
  Mechanics}}},\ \bibinfo {edition} {3rd}\ ed.,\ edited by\ \bibinfo {editor}
  {\bibfnamefont {R.}~\bibnamefont {Pathria}}\ and\ \bibinfo {editor}
  {\bibfnamefont {P.~D.}\ \bibnamefont {Beale}}\ (\bibinfo  {publisher}
  {Academic Press},\ \bibinfo {address} {Boston},\ \bibinfo {year}
  {2011})\BibitemShut {NoStop}%
\bibitem [{\citenamefont {Alder}\ and\ \citenamefont
  {Wainwright}(1970)}]{Alder:1970}%
  \BibitemOpen
  \bibfield  {author} {\bibinfo {author} {\bibfnamefont {B.~J.}\ \bibnamefont
  {Alder}}\ and\ \bibinfo {author} {\bibfnamefont {T.~E.}\ \bibnamefont
  {Wainwright}},\ }\href@noop {} {\bibfield  {journal} {\bibinfo  {journal}
  {Phys. Rev. A}\ }\textbf {\bibinfo {volume} {1}},\ \bibinfo {pages} {18}
  (\bibinfo {year} {1970})}\BibitemShut {NoStop}%
\bibitem [{\citenamefont {Landau}(1936)}]{Landau:1936}%
  \BibitemOpen
  \bibfield  {author} {\bibinfo {author} {\bibfnamefont {L.}~\bibnamefont
  {Landau}},\ }\href@noop {} {\bibfield  {journal} {\bibinfo  {journal} {Phys.
  Z. Sowjetunion}\ }\textbf {\bibinfo {volume} {10}},\ \bibinfo {pages} {154}
  (\bibinfo {year} {1936})}\BibitemShut {NoStop}%
\bibitem [{\citenamefont {Spitzer}\ and\ \citenamefont
  {H\"arm}(1953)}]{Spitzer:1953}%
  \BibitemOpen
  \bibfield  {author} {\bibinfo {author} {\bibfnamefont {L.}~\bibnamefont
  {Spitzer}}\ and\ \bibinfo {author} {\bibfnamefont {R.}~\bibnamefont
  {H\"arm}},\ }\href@noop {} {\bibfield  {journal} {\bibinfo  {journal} {Phys.
  Rev.}\ }\textbf {\bibinfo {volume} {89}},\ \bibinfo {pages} {977} (\bibinfo
  {year} {1953})}\BibitemShut {NoStop}%
\bibitem [{\citenamefont {Baalrud}\ and\ \citenamefont
  {Daligault}(2014)}]{Baalrud:EPT}%
  \BibitemOpen
  \bibfield  {author} {\bibinfo {author} {\bibfnamefont {S.~D.}\ \bibnamefont
  {Baalrud}}\ and\ \bibinfo {author} {\bibfnamefont {J.}~\bibnamefont
  {Daligault}},\ }\href@noop {} {\bibfield  {journal} {\bibinfo  {journal}
  {Phys. Plasmas}\ }\textbf {\bibinfo {volume} {21}},\ \bibinfo {pages}
  {055707} (\bibinfo {year} {2014})}\BibitemShut {NoStop}%
\bibitem [{\citenamefont {Chapman}\ and\ \citenamefont
  {Cowling}(1991)}]{Chapman:1991}%
  \BibitemOpen
  \bibfield  {author} {\bibinfo {author} {\bibfnamefont {S.}~\bibnamefont
  {Chapman}}\ and\ \bibinfo {author} {\bibfnamefont {T.~G.}\ \bibnamefont
  {Cowling}},\ }\href@noop {} {\emph {\bibinfo {title} {The Mathematical Theory
  of Non-Uniform Gases}}},\ \bibinfo {edition} {3rd}\ ed.\ (\bibinfo
  {publisher} {Cambridge University Press},\ \bibinfo {address} {Cambridge,
  UK},\ \bibinfo {year} {1991})\BibitemShut {NoStop}%
\bibitem [{\citenamefont {{Braginskii}}(1965)}]{Braginskii:1965}%
  \BibitemOpen
  \bibfield  {author} {\bibinfo {author} {\bibfnamefont {S.~I.}\ \bibnamefont
  {{Braginskii}}},\ }\href@noop {} {\bibfield  {journal} {\bibinfo  {journal}
  {Rev. Plasma Phys.}\ }\textbf {\bibinfo {volume} {vol. 1}} (\bibinfo {year}
  {1965})}\BibitemShut {NoStop}%
\bibitem [{\citenamefont {Dubin}(2014)}]{DubinPOP2014}%
  \BibitemOpen
  \bibfield  {author} {\bibinfo {author} {\bibfnamefont {D.~H.~E.}\
  \bibnamefont {Dubin}},\ }\href {\doibase 10.1063/1.4876749} {\bibfield
  {journal} {\bibinfo  {journal} {Physics of Plasmas}\ }\textbf {\bibinfo
  {volume} {21}},\ \bibinfo {pages} {052108} (\bibinfo {year} {2014})},\
  \Eprint {http://arxiv.org/abs/https://doi.org/10.1063/1.4876749}
  {https://doi.org/10.1063/1.4876749} \BibitemShut {NoStop}%
\bibitem [{\citenamefont {Jose}\ and\ \citenamefont
  {Baalrud}(2020)}]{Jose:2020}%
  \BibitemOpen
  \bibfield  {author} {\bibinfo {author} {\bibfnamefont {L.}~\bibnamefont
  {Jose}}\ and\ \bibinfo {author} {\bibfnamefont {S.~D.}\ \bibnamefont
  {Baalrud}},\ }\href@noop {} {\bibfield  {journal} {\bibinfo  {journal} {Phys.
  Plasmas}\ }\textbf {\bibinfo {volume} {27}},\ \bibinfo {pages} {112101}
  (\bibinfo {year} {2020})}\BibitemShut {NoStop}%
\bibitem [{\citenamefont {Stukowski}(2010)}]{ovito}%
  \BibitemOpen
  \bibfield  {author} {\bibinfo {author} {\bibfnamefont {A.}~\bibnamefont
  {Stukowski}},\ }\href@noop {} {\bibfield  {journal} {\bibinfo  {journal}
  {{Model. Simul. Mater. Sci. Eng.}}\ }\textbf {\bibinfo {volume} {{18}}}
  (\bibinfo {year} {{2010}})}\BibitemShut {NoStop}%
\bibitem [{\citenamefont {Lafleur}\ and\ \citenamefont
  {Baalrud}(2019)}]{Baalrud:Fx}%
  \BibitemOpen
  \bibfield  {author} {\bibinfo {author} {\bibfnamefont {T.}~\bibnamefont
  {Lafleur}}\ and\ \bibinfo {author} {\bibfnamefont {S.~D.}\ \bibnamefont
  {Baalrud}},\ }\href@noop {} {\bibfield  {journal} {\bibinfo  {journal}
  {Plasma Phys. Control. Fusion}\ }\textbf {\bibinfo {volume} {61}},\ \bibinfo
  {pages} {125004} (\bibinfo {year} {2019})}\BibitemShut {NoStop}%
\bibitem [{\citenamefont {Donk\'o}, \citenamefont {Kalman},\ and\ \citenamefont
  {Golden}(2002)}]{Donko:2002}%
  \BibitemOpen
  \bibfield  {author} {\bibinfo {author} {\bibfnamefont {Z.}~\bibnamefont
  {Donk\'o}}, \bibinfo {author} {\bibfnamefont {G.~J.}\ \bibnamefont {Kalman}},
  \ and\ \bibinfo {author} {\bibfnamefont {K.~I.}\ \bibnamefont {Golden}},\
  }\href@noop {} {\bibfield  {journal} {\bibinfo  {journal} {Phys. Rev. Lett.}\
  }\textbf {\bibinfo {volume} {88}},\ \bibinfo {pages} {225001} (\bibinfo
  {year} {2002})}\BibitemShut {NoStop}%
\bibitem [{\citenamefont {Ott}\ \emph {et~al.}(2012)\citenamefont {Ott},
  \citenamefont {K\"ahlert}, \citenamefont {Reynolds},\ and\ \citenamefont
  {Bonitz}}]{OttPRL2012}%
  \BibitemOpen
  \bibfield  {author} {\bibinfo {author} {\bibfnamefont {T.}~\bibnamefont
  {Ott}}, \bibinfo {author} {\bibfnamefont {H.}~\bibnamefont {K\"ahlert}},
  \bibinfo {author} {\bibfnamefont {A.}~\bibnamefont {Reynolds}}, \ and\
  \bibinfo {author} {\bibfnamefont {M.}~\bibnamefont {Bonitz}},\ }\href
  {\doibase 10.1103/PhysRevLett.108.255002} {\bibfield  {journal} {\bibinfo
  {journal} {Phys. Rev. Lett.}\ }\textbf {\bibinfo {volume} {108}},\ \bibinfo
  {pages} {255002} (\bibinfo {year} {2012})}\BibitemShut {NoStop}%
\bibitem [{\citenamefont {Ott}\ \emph {et~al.}(2013)\citenamefont {Ott},
  \citenamefont {Baiko}, \citenamefont {K\"ahlert},\ and\ \citenamefont
  {Bonitz}}]{OttPRE2013}%
  \BibitemOpen
  \bibfield  {author} {\bibinfo {author} {\bibfnamefont {T.}~\bibnamefont
  {Ott}}, \bibinfo {author} {\bibfnamefont {D.~A.}\ \bibnamefont {Baiko}},
  \bibinfo {author} {\bibfnamefont {H.}~\bibnamefont {K\"ahlert}}, \ and\
  \bibinfo {author} {\bibfnamefont {M.}~\bibnamefont {Bonitz}},\ }\href
  {\doibase 10.1103/PhysRevE.87.043102} {\bibfield  {journal} {\bibinfo
  {journal} {Phys. Rev. E}\ }\textbf {\bibinfo {volume} {87}},\ \bibinfo
  {pages} {043102} (\bibinfo {year} {2013})}\BibitemShut {NoStop}%
\bibitem [{\citenamefont {K\"{a}hlert}\ \emph {et~al.}(2013)\citenamefont
  {K\"{a}hlert}, \citenamefont {Ott}, \citenamefont {Reynolds}, \citenamefont
  {Kalman},\ and\ \citenamefont {Bonitz}}]{KahlertPOP2013}%
  \BibitemOpen
  \bibfield  {author} {\bibinfo {author} {\bibfnamefont {H.}~\bibnamefont
  {K\"{a}hlert}}, \bibinfo {author} {\bibfnamefont {T.}~\bibnamefont {Ott}},
  \bibinfo {author} {\bibfnamefont {A.}~\bibnamefont {Reynolds}}, \bibinfo
  {author} {\bibfnamefont {G.~J.}\ \bibnamefont {Kalman}}, \ and\ \bibinfo
  {author} {\bibfnamefont {M.}~\bibnamefont {Bonitz}},\ }\href {\doibase
  10.1063/1.4801522} {\bibfield  {journal} {\bibinfo  {journal} {Physics of
  Plasmas}\ }\textbf {\bibinfo {volume} {20}},\ \bibinfo {pages} {057301}
  (\bibinfo {year} {2013})},\ \Eprint
  {http://arxiv.org/abs/https://doi.org/10.1063/1.4801522}
  {https://doi.org/10.1063/1.4801522} \BibitemShut {NoStop}%
\bibitem [{\citenamefont {Hartmann}\ \emph {et~al.}(2013)\citenamefont
  {Hartmann}, \citenamefont {Donk\'o}, \citenamefont {Ott}, \citenamefont
  {K\"ahlert},\ and\ \citenamefont {Bonitz}}]{HartmannPRL2013}%
  \BibitemOpen
  \bibfield  {author} {\bibinfo {author} {\bibfnamefont {P.}~\bibnamefont
  {Hartmann}}, \bibinfo {author} {\bibfnamefont {Z.}~\bibnamefont {Donk\'o}},
  \bibinfo {author} {\bibfnamefont {T.}~\bibnamefont {Ott}}, \bibinfo {author}
  {\bibfnamefont {H.}~\bibnamefont {K\"ahlert}}, \ and\ \bibinfo {author}
  {\bibfnamefont {M.}~\bibnamefont {Bonitz}},\ }\href {\doibase
  10.1103/PhysRevLett.111.155002} {\bibfield  {journal} {\bibinfo  {journal}
  {Phys. Rev. Lett.}\ }\textbf {\bibinfo {volume} {111}},\ \bibinfo {pages}
  {155002} (\bibinfo {year} {2013})}\BibitemShut {NoStop}%
\bibitem [{\citenamefont {Dzhumagulova}\ \emph {et~al.}(2014)\citenamefont
  {Dzhumagulova}, \citenamefont {Masheeva}, \citenamefont {Ramazanov},\ and\
  \citenamefont {Donk\'o}}]{DzhumaulovaPRE2014}%
  \BibitemOpen
  \bibfield  {author} {\bibinfo {author} {\bibfnamefont {K.~N.}\ \bibnamefont
  {Dzhumagulova}}, \bibinfo {author} {\bibfnamefont {R.~U.}\ \bibnamefont
  {Masheeva}}, \bibinfo {author} {\bibfnamefont {T.~S.}\ \bibnamefont
  {Ramazanov}}, \ and\ \bibinfo {author} {\bibfnamefont {Z.}~\bibnamefont
  {Donk\'o}},\ }\href {\doibase 10.1103/PhysRevE.89.033104} {\bibfield
  {journal} {\bibinfo  {journal} {Phys. Rev. E}\ }\textbf {\bibinfo {volume}
  {89}},\ \bibinfo {pages} {033104} (\bibinfo {year} {2014})}\BibitemShut
  {NoStop}%
\bibitem [{\citenamefont {Towns}\ \emph {et~al.}(2014)\citenamefont {Towns},
  \citenamefont {Cockerill}, \citenamefont {Dahan}, \citenamefont {Foster},
  \citenamefont {Gaither}, \citenamefont {Grimshaw}, \citenamefont {Hazlewood},
  \citenamefont {Lathrop}, \citenamefont {Lifka}, \citenamefont {Peterson},
  \citenamefont {Roskies}, \citenamefont {Scott},\ and\ \citenamefont
  {Wilkins-Diehr}}]{xsede}%
  \BibitemOpen
  \bibfield  {author} {\bibinfo {author} {\bibfnamefont {J.}~\bibnamefont
  {Towns}}, \bibinfo {author} {\bibfnamefont {T.}~\bibnamefont {Cockerill}},
  \bibinfo {author} {\bibfnamefont {M.}~\bibnamefont {Dahan}}, \bibinfo
  {author} {\bibfnamefont {I.}~\bibnamefont {Foster}}, \bibinfo {author}
  {\bibfnamefont {K.}~\bibnamefont {Gaither}}, \bibinfo {author} {\bibfnamefont
  {A.}~\bibnamefont {Grimshaw}}, \bibinfo {author} {\bibfnamefont
  {V.}~\bibnamefont {Hazlewood}}, \bibinfo {author} {\bibfnamefont
  {S.}~\bibnamefont {Lathrop}}, \bibinfo {author} {\bibfnamefont
  {D.}~\bibnamefont {Lifka}}, \bibinfo {author} {\bibfnamefont {G.~D.}\
  \bibnamefont {Peterson}}, \bibinfo {author} {\bibfnamefont {R.}~\bibnamefont
  {Roskies}}, \bibinfo {author} {\bibfnamefont {J.~R.}\ \bibnamefont {Scott}},
  \ and\ \bibinfo {author} {\bibfnamefont {N.}~\bibnamefont {Wilkins-Diehr}},\
  }\href@noop {} {\bibfield  {journal} {\bibinfo  {journal} {Comput. Sci.
  Eng.}\ }\textbf {\bibinfo {volume} {16}},\ \bibinfo {pages} {62} (\bibinfo
  {year} {2014})}\BibitemShut {NoStop}%
\end{thebibliography}%

\end{document}